\documentclass[aip,pof,reprint]{revtex4-1}
\usepackage{bm}%
\usepackage{amsmath}
\usepackage{amssymb}
\usepackage{graphicx}
\usepackage{natbib}

\usepackage{dcolumn}
\usepackage[nooneline]{subfigure}

\expandafter\ifx\csname package@font\endcsname\relax\else
 \expandafter\expandafter
 \expandafter\usepackage
 \expandafter\expandafter
 \expandafter{\csname package@font\endcsname}%
\fi
\newcommand\Om{\hat{\boldsymbol{\Omega}}}
\newcommand\Li{\boldsymbol{\mathcal{L}}_i}
\newcommand\Lia{\boldsymbol{\mathcal{L}}_i^{*}}
\newcommand\Ji{\boldsymbol{\mathcal{J}}_i}
\newcommand\Jia{\boldsymbol{\mathcal{J}}_i^{*}}
\newcommand\Ui{\boldsymbol{\Upsilon}_i}
\newcommand\Pho{{\boldsymbol{\Phi}}_{1}^{i}}
\newcommand\Pht{{\boldsymbol{\Phi}}_{2}^{i}}
\newcommand\Pso{{\boldsymbol{\Psi}}_{1}^{i}}
\newcommand\Pst{{\boldsymbol{\Psi}}_{2}^{i}}
\newcommand\Pha{{\boldsymbol{\Phi}}_{a}^{i}}
\newcommand\Psa{{\boldsymbol{\Psi}}_{a}^{i}}
\newcommand\Rey{\mbox{\textit{Re}}}
\newcommand\Web{\mbox{\textit{W}\!\!\textit{e}}}

\begin{document}

\title{Rigorous coherent-structure theory for falling liquid films: Viscous dispersion effects on bound-state
formation and self-organization}
\author{Marc Pradas}
\affiliation{Department of Chemical Engineering, Imperial
College London, London, SW7 2AZ, United Kingdom}%
\author{Dmitri Tseluiko}
\affiliation{School of Mathematics, Loughborough University, Leicestershire,
LE11 3TU, United Kingdom}%
\author{Serafim Kalliadasis}
\affiliation{Department of Chemical Engineering, Imperial
College London, London, SW7 2AZ, United Kingdom}%
\date{\today}

\begin{abstract}
We examine the interaction of two-dimensional solitary pulses on
falling liquid films. We make use of the second-order model derived
by Ruyer-Quil and Manneville [Eur. Phys. J. B {\bf 6}, 277 (1998);
Eur. Phys. J. B {\bf 15}, 357 (2000); Phys. Fluids {\bf 14}, 170
(2002)] by combining the long-wave approximation with a weighted
residuals technique. The model includes (second-order) viscous
dispersion effects which originate from the streamwise momentum
equation and tangential stress balance. These effects play a
dispersive role that primarily influences the shape of the capillary
ripples in front of the solitary pulses. We show that different
physical parameters, such as surface tension and viscosity, play a
crucial role in the interaction between solitary pulses giving rise
eventually to the formation of bound states consisting of two or
more pulses separated by well-defined distances and travelling at
the same velocity. By developing a rigorous coherent-structure
theory, we are able to theoretically predict the pulse-separation
distances for which bound states are formed. Viscous dispersion
affects the distances at which bound states are observed. We show
that the theory is in very good agreement with computations of the
second-order model. We also demonstrate that the presence of bound
states allows the film free surface to reach a self-organized state
that can be statistically described in terms of a gas of solitary
waves separated by a typical mean distance and characterized by a
typical density.
\end{abstract}


\maketitle

\section{Introduction}

The dynamics of a liquid film falling down a vertical wall has been
the subject of numerous studies since the pioneering works by the
father-son Kapitza team~\cite{Kap48a,Kap48b,Kap49}. Wave evolution
on a falling film is a classical long-wave hydrodynamic instability
with a rich variety of spatial and temporal structures and a rich
spectrum of wave forms and wave transitions, starting from nearly
harmonic waves at the inlet to complex spatio-temporal highly
nonlinear wave patterns downstream. Such patterns are known to
profoundly affect the heat and mass transfer of multi-phase
industrial units. Reviews of the dynamics of a falling film are
given in Refs.~\onlinecite{Chang_ARFM_1994,Chang_Book_2002,KallTh07}.

For small-to-moderate values of the Reynolds number (defined
typically as the ratio of the inlet flow rate over the kinematic
viscosity), the falling-film surface is primarily dominated by
streamwise fluctuations and can be considered as free of spanwise
modulations, i.e. two dimensional \cite[][]{Dem07}. As it has been
observed in many experimental studies
\cite[][]{Krantz_IECF_1971,Gollub_PRL_1993,Liu_JFM_1993,Vlachogiannis_JFM_2001,Argyriadi_PhysFlui_2004}
and  theoretical works
\cite[][]{Pumir_JFM_1983,Chang_JFM_1995,Chang_JFM_2002}, under these
conditions, the film free surface
appears to be randomly covered by localized coherent structures,
each of which resembling (infinite-domain) solitary pulses. These
pulses are a consequence of a secondary modulation instability of
the primary wave field. They consist of a nonlinear hump preceded by
small capillary oscillations and can even appear at small Reynolds
numbers (but above the critical Reynolds number for the instability
onset which vanishes exactly for a vertical film).

Solitary pulses are stable and robust and continuously interact with
each other as quasi-particles through attractions and repulsions
giving rise to the formation of \emph{bound states} of two or more
pulses travelling at the same speed and separated by well-defined
distances. Bound-state formation of two or more pulses has been
recently observed experimentally in the problem of a viscous fluid
coating a vertical fiber~\cite[][]{Duprat_PRL_2009}. In this case,
the initial growth of the disturbances is driven by the
Rayleigh--Plateau instability and inertia. Eventually, the interface
is dominated by drop-like solitary pulses which continuously
interact with each other and can form bound states. The interplay
between, on the one hand the Rayleigh--Plateau instability and
inertia which enhance the front capillary ripples, and on the other
hand surface tension and viscous friction that suppress them, has a
crucial effect on the pulse interaction dynamics and, thus in turn,
on the distances at which bound states are formed.

In the present study we examine both analytically and numerically
the bound-state formation phenomena in a vertically falling liquid
film. We use a two-field model for the local flow rate and the
liquid free surface that contains terms up to second order in the
long-wave expansion
parameter~\cite[][]{Ruy98,RuyerQuil_EPJB_2000,Ruy02}. Hence, this
model includes the second-order viscous effects originating from the
streamwise momentum equation (streamwise viscous diffusion) and
tangential stress balance, i.e. second-order contributions to the
tangential stress at the free surface. These terms, which have been
ignored in all previous pulse interaction theories for film
flows~\cite[e.g.][]{Chang_JFM_1995,Chang_Book_2002}, have a
dispersive effect on the speed of the linear waves (they introduce a
wavenumber dependence on the speed) and they affect the shape of the
capillary ripples in front of a solitary hump. More specifically,
increasing the strength of the viscous dispersive effects leads to
decreasing the amplitude of the capillary waves ahead of the hump.
This effect is amplified as the Reynolds number is increased and
hence should play an important role on the selection process that
brings the pulses to be separated by well-defined distances.

We start by carefully developing a rigorous coherent-structure
theory for the second-order two-field model. The aim is to obtain a
dynamical system describing the location of each pulse by assuming
weak interaction between pulses (i.e. the pulses are sufficiently
far from each other and they interact through their tails only).
Although the basic ansatz we use at the outset, i.e. a superposition
of $N$ pulses plus an overlap function, is a standard assumption in
weak interaction theories (and originates from condensed matter
physics where it has been used to describe particle-particle
interaction), the way we implement it into the two-field model is
highly non trivial. For example, the spectral analysis of the
resulting linearized non self-adjoint operator describing soliton
interaction requires a careful and rigorous study that has not been
performed before for a two-field system. For a single equation, the
generalized Kuramoto--Sivashinsky (gKS) equation, a careful and
rigorous study was performed recently in
Refs.~\onlinecite{Duprat_PRL_2009,Tse10,Tseluiko_PD_2010}. Here we
appropriately extend this study to the second-order two-field model.

Of course, coherent-structure theories have been formulated for many
different systems in recent years (see
e.g.~Ref.~\onlinecite{Balmforth1995} for the review of some of the
methodologies for the gKS equation). Even rigorous justification of
the ordinary-differential equations describing the dynamics of the
pulses has been provided recently for equations with a stable
primary pulse (e.g.~Refs.~\onlinecite{Ei2002,Zelik2009}). However,
for the present problem, as well as for the gKS equation analyzed
recently in
Refs.~\onlinecite{Duprat_PRL_2009,Tse10,Tseluiko_PD_2010}, the
pulses are inherently unstable with the zero eigenvalue of the
linearized interaction operator embedded into the essential
spectrum. This in turn suggests that the usual projection procedure
used in previous studies, such as those on weak-interaction
approaches for the gKS equation
(e.g.~Refs.~\onlinecite{Elphick1991,EiOhta1994,Chang_Book_2002}),
cannot be rigourously justified. Moreover, previous studies appear
to be either incomplete or at times overlook important details and
subtleties. For example, the structure of the spectra of the adjoint
operator of the linearized equation for the overlap function in the
vicinity of a pulse, a crucial step for performing projections, has
not been analyzed carefully. This is done here by considering the
linearized interaction operator on a finite domain and by imposing
periodic boundary conditions. In this way we are able to recover the
eigenvalues and the corresponding eigenfunctions on an infinite
interval in the limit of the periodicity interval tending to
infinity. We then show that the null adjoint eigenfunction has a
jump at infinity, which in turn implies that the localized function
in the null space of the adjoint operator given in
Ref.~\onlinecite{Chang_Book_2002} (Fig.~9.1(c)) and also postulated
in Ref.~\onlinecite{EiOhta1994} is erroneous (these misconceptions
are also discussed in the recent study by Tseluiko \emph{et
al.}~\cite{Tseluiko_PD_2010}).

As we shall also demonstrate here the projections for the derivation
of the system governing the pulse dynamics can be made rigorous by
the use of weighted functional spaces. We are then able to obtain
rigorously a dynamical system describing the time evolution for the
pulse locations. All possible distances at which bound states are
formed are found by investigating the fixed points of this dynamical
system. Detailed statistical analysis of time-dependent computations
with the second-order model then shows that the separation distances
between neighboring structures that the system selects are in
excellent agreement with those predicted by the coherent-structure
theory. Moreover, the time-dependent computations with the
second-order model elucidate the influence of viscous dispersion:
increasing viscous dispersion allows the pulses to get closer to
each other thus decreasing the separation distances between
neighboring coherent structures.

The use of weighted spaces makes our solitary pulses spectrally
stable and hence allows also for the derivation of a novel and
highly effective numerical scheme that can be used to accurately
track the pulse dynamics for sufficiently long times. Furthermore,
even though we primarily focus on weak interaction between solitary
pulses, we also give numerical results on the strong interaction
case, i.e. when the pulses are sufficiently close to each other,
revealing a peculiar oscillatory behavior. Although an appropriate
strong interaction theory describing this phenomenon is still
lacking, we show that considering or not the second-order viscous
dispersion effects becomes crucial for the description of the strong
interaction dynamics.

In Sec.~\ref{Sec:Model} we present the second-order model that
includes the viscous-dispersion effects. In Sec.~\ref{Sec:Theory} we
develop a coherent-structure theory for the interaction of solitary
pulses of the second-order model. In Sec.~\ref{Sec:num} we compare
our theoretical predictions for formation of bound states with
time-dependent computations of the full system. Finally, we conclude
in Sec.~\ref{Sec:concl}.

\section{Second-order model}\label{Sec:Model}
\subsection{General formulation} \label{Sec:General Formulation}

\begin{figure}
\centering
\includegraphics[width=0.16\textwidth]{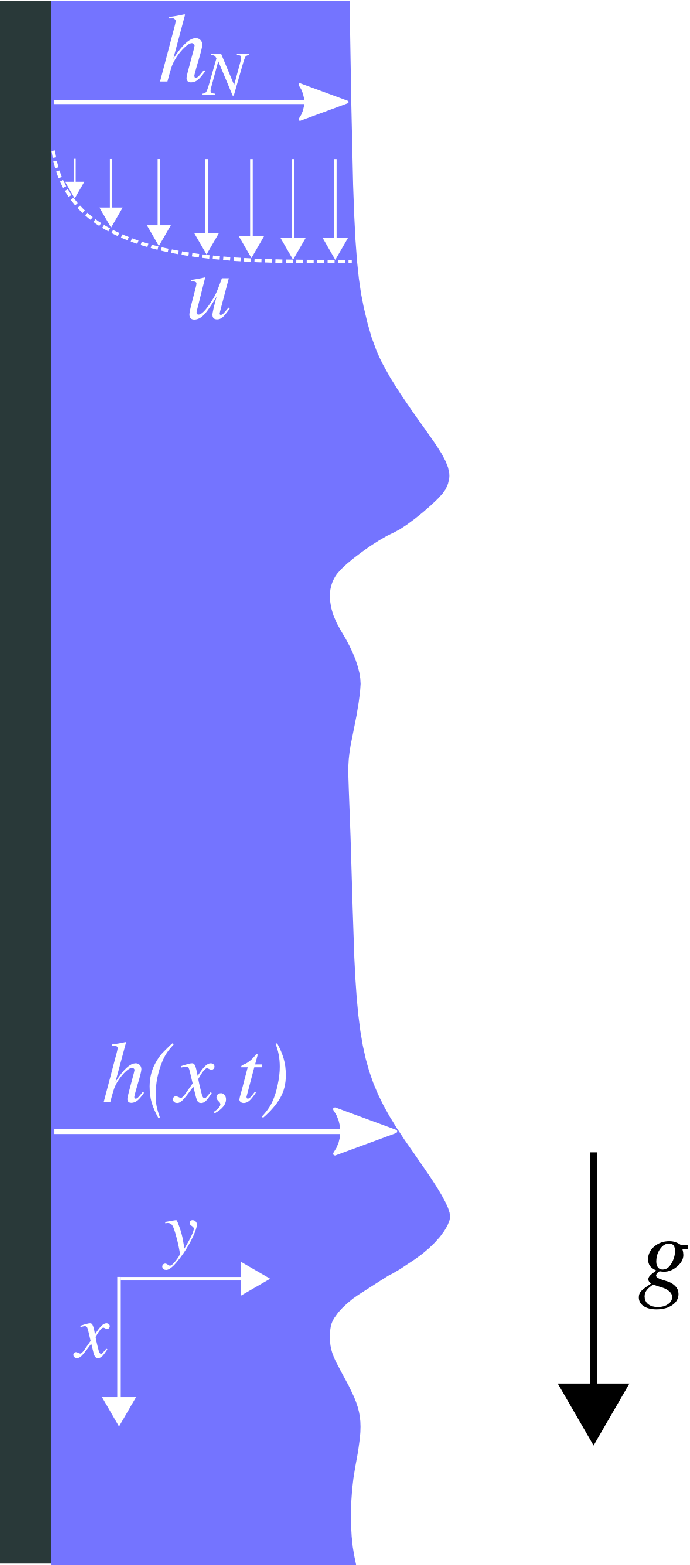}\\
\caption{(Color online) Sketch of the profile geometry for a two-dimensional liquid
film flowing under the action of gravity down a vertical
wall.}\label{Fig:setup}
\end{figure}

Figure~\ref{Fig:setup} shows the problem definition. We consider a
thin liquid film flowing under the action of gravity down a vertical
planar substrate. The liquid has density $\rho$, kinematic viscosity
$\nu$ and surface tension $\sigma$. A Cartesian coordinate system
$(x,y)$ is introduced so that $x$ is in the direction parallel to
the wall and $y$ is the outward-pointing coordinate normal to the
wall. The wall is then located at $y=0$ and the free surface at
$y=h(x,t)$. The governing equations are the mass conservation and
Navier--Stokes equations along with the no-slip and no-penetration
conditions on the wall and the kinematic and tangential and normal
stress balance conditions on the free surface.

By introducing a formal parameter $\epsilon$ representing a typical
slope of the film, $\epsilon \sim |\partial_x h|/h$, we can perform
a long-wave expansion of the equations of motion and associated wall
and free-surface boundary conditions for $\epsilon \ll 1$. This is
based on the observation that because surface tension is generally
large, the interfacial waves are typically long compared to the film
thickness. This so-called ``long-wave approximation" has been
central to all thin-film studies (see e.g.
Refs.~\onlinecite{ODB97,KallTh07})
and the small parameter $\epsilon$ is
frequently referred to as the ``long-wave" or ``film parameter".
The long-wave approximation leads to a hierarchy of model equations,
starting from a single highly nonlinear partial differential
equation for the film thickness $h$, the so-called ``Benney
equation"~\cite[][]{JMathPhys_45_150} whose region of validity is
restricted to near-critical conditions, to equations of the
boundary-layer type which are valid away from
criticality~\cite[]{Chang_ARFM_1994,Chang_Book_2002,KallTh07}. By
neglecting terms of $O(\epsilon^3)$ and higher, the so-called
``second-order boundary-layer equations" read:
\begin{subequations}
\label{Eq:BL NS}%
\begin{eqnarray}
u_x+v_y & = & 0, \label{Eq:BL NS a}{} \\
\Rey(u_t+uu_x+vu_y) & = &
 3(1+\Web h_{xxx}) \nonumber {} \\
& & +u_{yy}-2u_{xx}+[u_x|_h]_x, \label{Eq:BL NS b} {} \\
u|_0 & = & v|_0=0, \label{Eq:BL NS c}{} \\
u_y|_h & = & 4h_x(u_x|_h)-v_x|_h, \label{Eq:BL NS d}{} \\
h_t+q_x & = & 0, \label{Eq:BL NS e}
\end{eqnarray}
\end{subequations}
where $x$ and $y$ have been non-dimensionalized with the Nusselt
flat-film thickness $h_N$, the streamwise and cross-stream velocity
components, $u$ and $v$, respectively, are both non-dimensionalized
by the average Nusselt flat-film velocity, $u_N
=\mathrm{g}h_N^2/3\nu$, where $\mathrm g$ is the gravitational
acceleration, and time $t$ is non-dimensionalized by $h_N/u_N$.
Here, $q=\int_0^hu\:\mathrm{d}y$ is the streamwise flow rate while
the notations $(\cdot)|_0$ and $(\cdot)|_h$ indicate that the
corresponding function is evaluated at $y=0$ and $y=h(x,t)$,
respectively. The two dimensionless groups appearing in~(\ref{Eq:BL
NS b}) are the Reynolds number, measuring the relative importance of
inertia to viscous forces and the Weber number, measuring the
relative importance of surface tension to gravity, defined as:
\begin{equation}
\Rey=\frac{u_N h_N}{\nu},\qquad \Web=\frac{\sigma}{\rho \mathrm{g}
h_N^2}.
\end{equation} We note that elimination of the pressure from the
cross-stream momentum equation constitutes the main element of the
boundary-layer approximation: by neglecting the inertia terms in
this equation, the resulting equation can be integrated across the
film to yield the pressure distribution in the film which in turn is
substituted into the streamwise momentum equation. It is also
important to note that the second-order contributions in the
long-wave expansion are given by the two last terms in the
right-hand sides of (\ref{Eq:BL NS b}) and (\ref{Eq:BL NS d})
\cite[]{RuyerQuil_EPJB_2000}. Hence, the second-order boundary layer
equations include streamwise viscous-diffusion effects and
second-order contributions to the tangential stress at the free
surface. These terms play a dispersive role, i.e. they introduce a
wavenumber dependence on the speed of the linear waves~\cite{Ruy02}.

By combining the long-wave expansion with a weighted residuals
technique based on Galerkin projection in which the velocity field
is expanded onto a basis with polynomial test functions, Ruyer-Quil
and Manneville~\cite{Ruy98,RuyerQuil_EPJB_2000,Ruy02} obtained the
following second-order two-field model for the local film thickness
and flow rate:
\begin{subequations}
\label{Eq:Model}%
\begin{eqnarray}
\delta q_t & = & \frac{5}{6}h - \frac{5}{2} \frac{q}{h^2} +
\delta\bigg(\frac97\frac{q^2}{h^2}h_x -
\frac{17}{7}\frac{q}{h}q_x\bigg)+ \frac{5}{6} h h_{xxx} \nonumber {} \\
& + & \eta\bigg[4\frac{q}{h^2}(h_x)^2-\frac{9}{2h}q_x
h_x-6\frac{q}{h}h_{xx}+\frac92q_{xx}\bigg], \label{Eq:Model a}{} \\
& & \nonumber {} \\
h_t &= & -q_x, \label{Eq:Model b}
\end{eqnarray}
\end{subequations}
where lengths, time and velocities in~(\ref{Eq:BL NS})
have been rescaled using the following scaling due to Shkadov~\cite{Shk77}
\begin{eqnarray}
(x,y) & \mapsto & (\kappa x,y), {} \\
(u,v) & \mapsto & (u,\kappa^{-1}v), {} \\
t     & \mapsto & \kappa t,
\end{eqnarray}
where $\kappa = \Web^{1/3}$ and
\begin{equation}
 \delta = \frac{3\Rey}{\kappa}, \qquad \eta = \frac{1}{\kappa^2},
\end{equation}
corresponding to a \emph{reduced Reynolds} and
\emph{viscous-dispersion number}, respectively (note that all
second-order/viscous-dispersion terms are gathered in the second
line of~(\ref{Eq:Model a})). For $\eta=0$ we obtain the first-order
model \cite[][]{RuyerQuil_EPJB_2000}, which, much like the
second-order one, can be derived from the first-order boundary-layer
equations using the long-wave expansion and a weighted residuals
technique. It should also be noted that the second-order model contains the
same number of parameters as the second-order boundary-layer
equations in~(\ref{Eq:BL NS}) and hence this model retains the
``structure" of these equations (in contrast e.g. with the boundary-layer
theory of aerodynamics where the Reynolds number can be scaled away
from the boundary layer as the corresponding equations are
``simpler" compared to full Navier--Stokes).

This first-order model is similar to the first-order model obtained
by Shkadov~\cite{IzvAkadNaukSSSR_1_43} by combining the long-wave
approximation and  averaging the basic equations across the fluid
layer (effectively, a weighted residual technique with weight
function equal to unity) but with different coefficients. As was
pointed by Ruyer-Quil and Manneville~\cite{RuyerQuil_EPJB_2000} the
second-order model in~(\ref{Eq:Model}) corrects the shortcomings of
the first-order model obtained by
Shkadov~\cite{IzvAkadNaukSSSR_1_43}, the principal one being
erroneous prediction of the instability onset, i.e. of critical and
neutral quantities: (a) it yields a 20$\%$ error of the critical
Reynolds number for an inclined film (once again, for a vertical
plane the critical Reynolds number vanishes) and (b) the neutral
curve (wavenumber for the instability as a function of the Reynolds
number) is not in agreement with Orr--Sommerfeld; for this purpose
the second-order viscous-dispersion terms are crucial. This is a
point that was analyzed carefully by Ruyer-Quil and
Manneville~\cite{Ruy98}. These authors contrasted the neutral curve
obtained from both the first- and second-order model to the one
obtained from the Orr--Sommerfeld eigenvalue problem. The comparison
shows clearly that the second-order model is in better agreement
with Orr--Sommerfeld than the first-order one.

Finally, it is noteworthy that the parameters $\delta$ and $\eta$ can
be expressed as
\begin{equation}
\delta=\Gamma^{-1/3}(3\Rey)^{11/9},
\qquad
\eta=\Gamma^{-2/3}(3\Rey)^{4/9},
\end{equation}
where $\Gamma=\sigma/(\rho g^{1/3}\nu^{4/3})$ is the Kapitza number
that depends on the physical properties of the liquid only. It can
then be readily seen that the second-order contributions, controlled
by $\eta$, are expected to be more relevant as $\Rey$ increases and
for liquids with either small surface tension or large viscosity.

\subsection{Steady-state solutions: solitary pulses}
\label{Sec:Model steady}

Solitary pulses are traveling-wave solutions propagating at constant
speed $c_0\equiv c_0(\delta,\eta)$, hence stationary in a frame
moving with speed $c_0$, and sufficiently localized in space.
Introducing then in (\ref{Eq:Model}) the moving coordinate $x\to
x-c_0 t$ and requiring that waves are stationary in this moving
frame, yields:
\begin{subequations}
\label{Eq:Model Steady}%
\begin{eqnarray}
& & c_0 q_{0x}-\frac{q_0}{7h_0}\bigg (17q_{0x}-
\frac{9q_0}{h_0}h_{0x}\bigg)+\frac{5h_0}{6\delta}-\frac{5q_0}{2\delta h_0^2} \nonumber {} \\
& &  +\frac{5h_0}{6\delta}h_{0xxx}
+\frac{\eta}{\delta}\bigg[\frac{4q_0}{h_0^2}(h_{0x})^2-
\frac{9}{2h_0}q_{0x}h_{0x} \nonumber {} \\
& &-\frac{6q_0}{h_0} h_{0xx}+\frac92q_{0xx}\bigg]=0, \label{Eq:Model Steady a}{} \\
& & \nonumber {} \\
&& c_0 h_{0x}-q_{0x}=0, \label{Eq:Model Steady b}
\end{eqnarray}
\end{subequations}
where $q_0(x)$ and $h_0(x)$ are the stationary solutions for the
local flow rate and free-surface shape, respectively. Integrating
once~(\ref{Eq:Model Steady}$b$) yields $q_0=c_0 h_0+\alpha$, where
$\alpha$ is an integration constant that can be fixed by demanding
that the free-surface height approaches the Nusselt flat film
solution away from the solitary hump, i.e. $h_0=1$ as $x \to \pm
\infty$, giving
\begin{equation}\
 q_0(x)=c_0[h_0(x)-1]+\frac{1}{3},
\end{equation}
which in turn is used to eliminate $q_0$ in~(\ref{Eq:Model
Steady}$a$). The resulting equation is solved numerically with a
pseudo-spectral scheme in a periodic domain $[-L,L]$ in which we
take the Fast Fourier Transform (FFT) of the equation to obtain a
system of nonlinear algebraic equations for the unknown FFT
components of $h_0$ and the pulse speed $c_0$. This system is solved
using Newton's method by choosing an appropriate initial guess.
\begin{figure}
\centering
\includegraphics[width=0.48\textwidth]{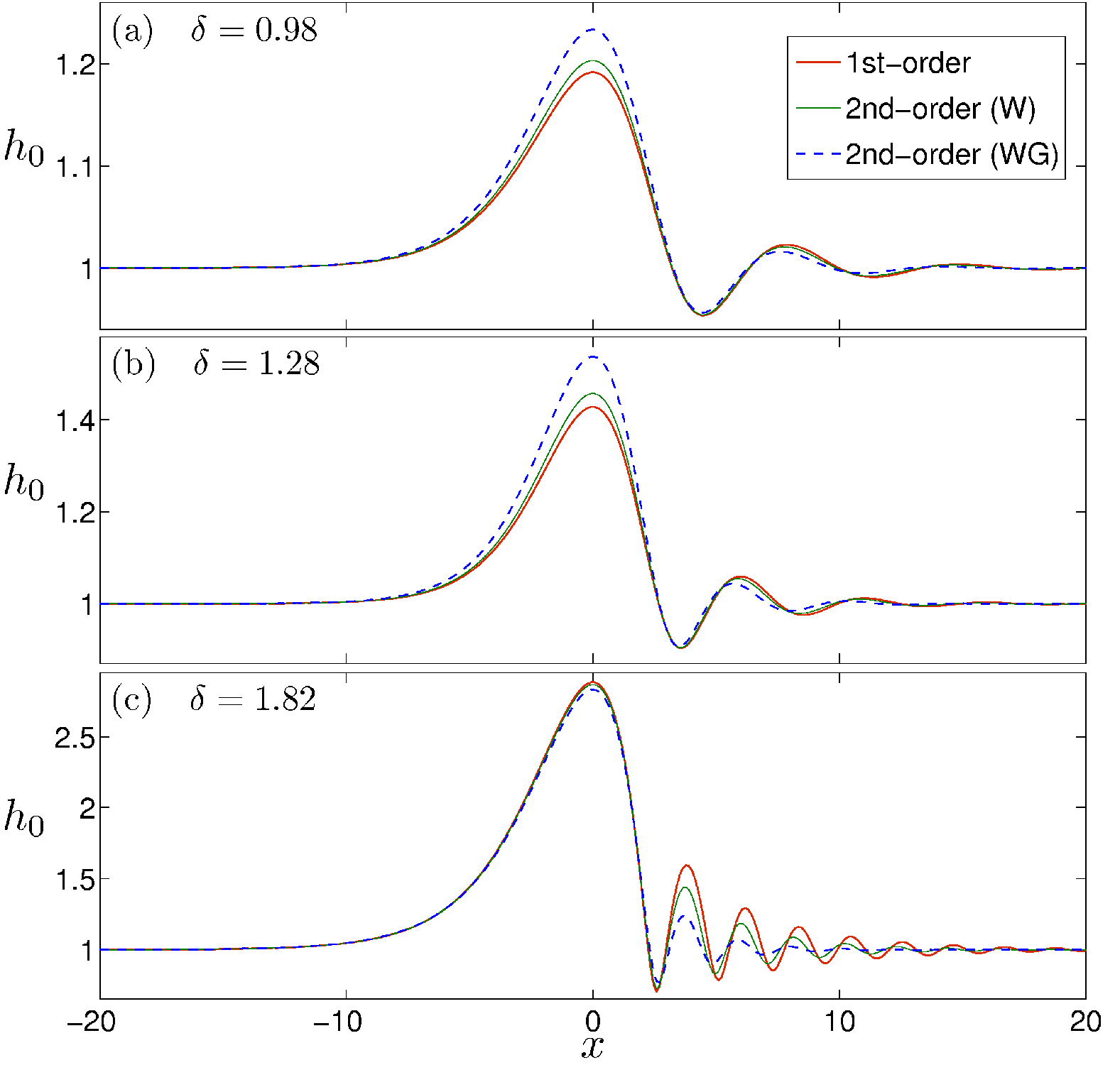}\\
\caption{(Color online) Stationary pulse profiles for different values of $\delta$
and two different liquids: water (W) and water+glycerin (WG) for
both the first-order ($\eta=0$) and the second-order ($\eta>0$)
models.}\label{Fig:steady}
\end{figure}

Figure \ref{Fig:steady} shows several examples of stationary
profiles $h_0(x)$. We have used the values $\delta=0.98$, $1.28$,
and $1.82$, which correspond to water (W) at $\Rey=3$, $3.75$, and
$5$, respectively, and to a water and glycerin 50$\%$ by weight
mixture (WG) at $\Rey=1.59$, $2.0$, and $2.66$, respectively (both
were used as working fluids in the falling film experiments carried
out by Liu \emph{et al.}~\cite{Liu_JFM_1993}, but for inclined films
with small inclination angle).
The kinematic
viscosities of W and WG are $\nu_W=10^{-6}$ m$^2$/s and
$\nu_{WG}=5\nu_W$, respectively, the densities are $\rho_W=1.0$
g/cm$^3$ and $\rho_{WG}=1.13$ g/cm$^3$, respectively, the surface
tensions are $\sigma_W=69\times 10^{-3}$ N/m and $\sigma_{WG}=72\times 10^{-3}$
N/m, respectively, and the resulting Kapitza numbers are $\Gamma_W=3364$ and
$\Gamma_{WG}=334$, respectively.
It is worth noting that as long
as $\delta$ is kept fixed, the first-order model ($\eta=0$) cannot
distinguish, for instance, between W at $\Rey=3$ and WG at
$\Rey=1.59$. Therefore, the results for the first-order model
correspond to both physical situations.

For $\delta=0.98$, $1.28$, and $1.82$, we obtain the following
values of the pulse speed and the maximum height, ($c_0,h_m$):
($1.15,1.19$), ($1.34,1.43$), and ($2.44,2.88$), respectively, from
the first-order model, and ($1.15,1,20$), ($1.36,1.46$), and
($2.43,2.86$), respectively, from the second-order model using W,
and ($1.18,1.23$), ($1.42,1.53$), and ($2.4,2.83$), respectively,
from the second-order model using WG. As it has been noticed by
Ruyer-Quil and Manneville~\cite{RuyerQuil_EPJB_2000}, the
differences between the first- and the second-order models as far as
solitary pulses are concerned become more significant as $\delta$ is
increased. Although the pulse speed and the maximum height do not
differ much (differences are appreciable at low values of $\delta$),
the capillary ripples that are present downstream of the hump appear
to be largely suppressed as a consequence of the second-order
viscous-dispersion effects, specially for WG which has higher viscosity.


\section{Coherent-structure theory for interacting pulses}
\label{Sec:Theory}

As it has been emphasized in the Introduction, at sufficiently large
distances from the inlet of the film, the dynamics of the free
surface is dominated by the presence of localized coherent
structures, each of which resembling (infinite-domain) solitary
pulses, which continuously interact with each other as quasi
particles through attractions and repulsions. The objective of this
section is to appropriately extend the recently developed
coherent-structure interaction theory for the solitary pulses of the
gKS equation~\cite[][]{Duprat_PRL_2009,Tse10,Tseluiko_PD_2010} to
the two-field model given by (\ref{Eq:Model}). This is by far a
non-trivial task as we shall demonstrate, e.g.~unlike the scalar gKS
equation~\cite{Duprat_PRL_2009,Tse10,Tseluiko_PD_2010} we now deal
with a rather involved vector equation. The
aim is to obtain a dynamical system describing the location of each
pulse by assuming weak interaction between pulses (i.e.~the pulses
are sufficiently far from each other and they interact through their
tails only). This concept has been applied in many other fields,
such as particle physics and quantum mechanics, where one typically
deals with a system compound of many particles, and has been used
successfully to describe particle--particle interaction. As
emphasized in the Introduction, previous coherent-structure theories
appear to be either incomplete or some times overlook important
details and subtleties, for instance in relation of the spectrum of
the operator describing interaction. Moreover, previous
coherent-structure theories for film flows in the region of
small-to-moderate Reynolds
numbers~\cite[e.g.][]{Chang_JFM_1995,Chang_Book_2002} were based on
the Shkadov model and thus ignored the effect of viscous dispersion.

\begin{figure}
\centering
\includegraphics[width=0.48\textwidth]{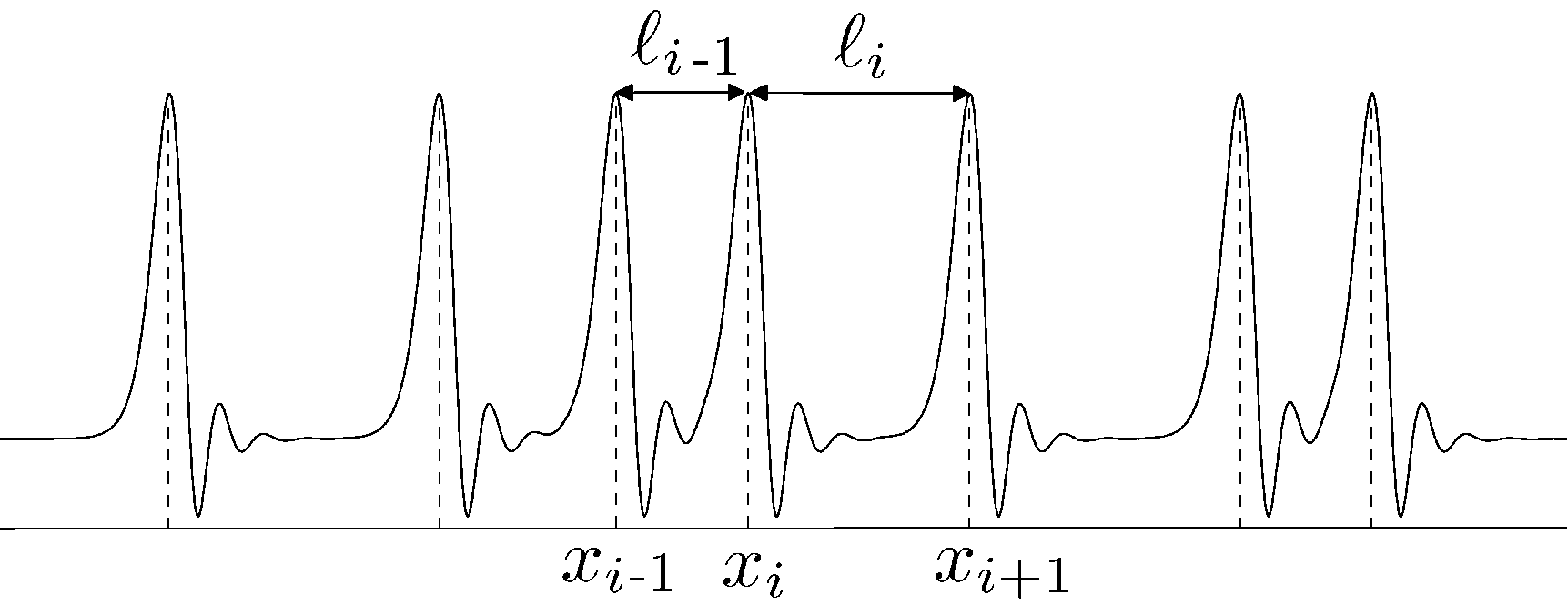}\\
\caption{Schematic representation of a solution consisting
of a supersposition of $N$ pulses located at $x_{i}$
for $i=1,\,\ldots,\,N$.}\label{Fig:N pulses}
\end{figure}

We start by considering the system of partial differential
equations~(\ref{Eq:Model}) in the frame moving with velocity $c_0$
of a single stationary pulse, $x\to x-c_0 t$. Then we assume that a
solution for the local flow rate, $q(x,t)$, and free-surface
profile, $h(x,t)$, is given as a superposition of $N$
quasi-stationary pulses located at $x_1(t),\,\ldots,\,x_N(t)$
(the pulses are labeled from left to right, so that $x_1<\cdots<x_N$) and a
small overlap function, i.e., we postulate the following ansatz:
\begin{subequations}
\label{Eq:Ansatz}
\begin{eqnarray}
q(x,t)&=&\frac13+\sum_{i=1}^{N}Q_i(x,t)+\hat{Q}(x,t),\label{Eq:Ansatz a} {} \\
h(x,t)&=&1+\sum_{i=1}^{N}H_i(x,t)+\hat{H}(x,t),\label{Eq:Ansatz b}
\end{eqnarray}
\end{subequations}
where $Q_i(x,t)=q_i(x,t)-1/3$ and  $H_i(x,t)=h_i(x,t)-1$ and
\begin{equation}
q_i(x,t)\equiv q_0(x-x_i(t))\qquad h_i(x,t)\equiv
h_0(x-x_i(t)),
\end{equation}
with $q_0$ and $h_0$ denoting the steady-state
solution of (\ref{Eq:Model Steady}).
A schematic representation of an $N$-pulse solution is given in
Fig.~\ref{Fig:N pulses}, where we have defined the distances between
two pulses as $\ell_i=x_{i+1}-x_{i}$, for  $i=1,\,\ldots,\,N-1$. It
is worth to mention here that such general formalism derived in the
following for the case of $N$ pulses will be ultimately applied to a
more simplified system of two pulses only. Next, we consider weak
interaction between the pulses by assuming that they are well
separated, i.e. $\ell_i\gg 1$,  so that for each pulse it is enough
to consider its interaction with only the immediate neighbours.
Because of the exponential decay of functions $H_0(x)=h_0(x)-1$ and
$Q_0(x)=q_0(x)-1/3$ as $x\rightarrow\pm\infty$, there exist positive
constants $C$ and $a$ such that $|H_0(x)|\leq C\exp(-a|x|)$ and
$|Q_0(x)|\leq C\exp(-a|x|)$. We define a small parameter
$\varepsilon=\exp{(-a\min_i\{\ell_i\})}$. Then in the vicinity of
each pulse $i$, the tails of the pulses $i-1$ and $i+1$ are
$O(\varepsilon)$ and the tails of the remaining pulses are
$o(\varepsilon)$.
We also assume that the pulse velocities $\dot{x}_i(t)$ and the
overlap functions $\hat{Q}$ and $\hat{H}$ are $O(\varepsilon)$.
By substituting (\ref{Eq:Ansatz}) into (\ref{Eq:Model}) and
expanding up to $O(\varepsilon)$, we obtain a linearized equation in
the vicinity of the $i$th pulse for the overlap vector function
\begin{equation}
\Om = \binom{\hat{Q}}{\hat{H}},
\end{equation}
that is written as follows:
\begin{equation}\label{Eq:Linear}
\Om_t -\dot{x}_i(t){\boldsymbol{\Phi}}_{ix}=\Li
\Om + \Ji \Ui,
\end{equation}
where
\begin{equation}\label{Eq: Defs}
{\boldsymbol{\Phi}}_{i}= \binom{q_{i}}{h_{i}},\qquad\boldsymbol{\Upsilon}_i = \binom{Q_{i-1}+Q_{i+1}}{H_{i-1}+
H_{i+1}},
\end{equation}
for $i=1,\,\ldots,\,N$. Note that for $i=1$ and $i=N$, the vector
$\boldsymbol{\Upsilon}$ reads as: $\boldsymbol{\Upsilon}_1 =
(Q_{2},H_{2})^T$ and $\boldsymbol{\Upsilon}_N =
(Q_{N-1},H_{N-1})^T$. $\Li$ and $\Ji$ are the following linear
matrix/differential operators:
\begin{equation}
 \Li=\begin{pmatrix}
  \mathcal{L}_i^{1} & \mathcal{L}_i^{2} \\
   \mathcal{L}_i^{3}& \mathcal{L}_i^{4}
\end{pmatrix},\qquad \Ji=\begin{pmatrix}
  \mathcal{J}_i^{1} & \mathcal{J}_i^{2} \\
   0 & 0
\end{pmatrix},
\end{equation}
with the components
\begin{eqnarray*}
\mathcal{L}_i^1 & = & -\frac{5}{2\delta}\frac{1}{h_i^2}+\frac{18}{7}
\frac{q_i}{h_i^2}h_{ix}-\frac{17}{7}\frac{q_{ix}}{h_i}+\bigg(c_0-
\frac{17}{7}\frac{q_i}{h_i}\bigg)\partial_x \nonumber {} \\
& & +\frac{\eta}{\delta}\bigg(4\frac{h_{ix}^2}{h_i^2}-6
\frac{h_{ixx}}{h_i}-\frac92\frac{h_{ix}}{h_i}\partial_x+
\frac92\partial_{xx}\bigg),  {} \\
\mathcal{L}_i^2& = & \frac{5}{6\delta}+\frac{5}{\delta}\frac{q_i}{h_i^3}+
\frac{17}{7}q_i\frac{q_{ix}}{h_i^2}-\frac{18}{7}q_i^2
\frac{h_{ix}}{h_i^3}+\frac{5}{6\delta}h_{ixxx} {} \\
& & + \frac97\frac{q_i^2}{h_i^2}\partial_x +\frac{5}{6\delta}h_i
\partial_{xxx}+\frac{\eta}{\delta}\bigg[\frac92 q_{ix}\frac{h_{ix}}{h_i^2}-
8q_i\frac{h_{ix}^2}{h_i^3} \nonumber {} \\
& & +6q_i\frac{h_{ixx}}{h_i^2}+\bigg(8q_i\frac{h_{ix}}{h_i^2}
-\frac92\frac{q_{ix}}{h_i}\bigg)\partial_x-
6\frac{q_i}{h_i}\partial_{xx} \bigg], {} \\
\mathcal{L}_i^3 & = & -\partial_x,  {} \\
\mathcal{L}_i^4 & = & c_0\partial_x,
\end{eqnarray*}
and
\begin{eqnarray*}
\mathcal{J}_i^{1} & = & \frac{18q_i}{7h_i^2}H_{ix}-\frac{17}{7 h_i}(Q_{ix}+
Q_i\partial_x)\nonumber  {} \\
& &+\frac{\eta}{\delta}\bigg(\frac{4}{h_i^2}H_{ix}^2-
\frac{6}{h_i}H_{ixx}-\frac{9}{2 h_i}H_{ix}\partial_x\bigg),
\nonumber  {} \\
\mathcal{J}_i^{2} & = & \frac{5}{\delta h_i^4}(q_iH_i+Q_i)+
\frac{17q_i}{7h_i^2}Q_{ix}-\frac{18q_i^2}{7h_i^3}H_{ix} {} \\
& & + \frac{9q_i^2-1}{7h_i^2}\partial_x+\frac{5}{6\delta}(H_{ixxx}+
H_i\partial_{xxx})\nonumber {} \\
& & +\frac{\eta}{\delta}\bigg[\frac{9}{2 h_i^2} Q_{ix}H_{ix}-
\frac{8q_i}{h_i^3}H_{ix}^2+\frac{6q_i}{h_i^2}H_{ixx} {} \\
& &+\bigg(\frac{8q_i}{h_i^2}H_{ix}-\frac{9}{2 h_i}Q_{ix}\bigg)
\partial_x-\frac{6}{h_i}Q_i\partial_{xx} \bigg]. \nonumber
\end{eqnarray*}
Equation (\ref{Eq:Linear}) reveals that the dynamics of the overlap
function in the vicinity of the $i$th pulse depends on the spectrum
of the linear operator $\Li$ and is influenced by the neighbouring
pulses, as indicated by the last term on the right-hand side of
(\ref{Eq:Linear}).

\subsection{Analysis of the structure of spectrum of the linearized operator describing
interaction}

It can be verified numerically that on a periodic domain the
operator $\Li$ has a zero eigenvalue with geometric multiplicity one
and algebraic multiplicity two (the numerical scheme for
constructing the spectrum of $\Li$ will be described shortly). The
operator $\Li$ then has a null space spanned by the eigenfunction
$\Pho\equiv{\boldsymbol{\Phi}}_{ix}$ that is associated with the
translational invariance of the system. The corresponding
generalized zero eigenfunction $\Pht$ that satisfies $\Li\Pht=\Pho$
is associated with the Galilean invariance of the system. The aim
then is to project the dynamics of the overlap function onto the
null space of $\Li$ in the vicinity of the $i$th pulse.

From a physical point of view, the existence of these two dominant
modes means that any perturbation to the steady pulse solution will
make the pulse to shift (due to the translational mode) and/or to
accelerate (due to the Galilean mode). We note that according to the
solution ansatz (\ref{Eq:Ansatz}), we can assume that the overlap
function, $\Om$, is ``free of translational modes". The precise
meaning of the latter phrase will be explained later.

Projection onto the null space of the linear matrix/differential
operator $\Li$ requires a careful and detailed analysis of its
spectrum as well as the spectrum of the adjoint operator $\Lia$ (see
Appendix~\ref{AppA}). The essential spectrum of the operator $\Li$
is given by the dispersion relation of the basic Nusselt state,
$(q_N,h_N)=(1/3,1)$, in the moving frame. By replacing $(q_i,h_i)$
in $\Li$ with the Nusselt state, $\Li$ becomes a matrix operator
with constant coefficients and its essential spectrum $\lambda(k)$
satisfies the following equation:
\begin{widetext}
\begin{equation}\label{Eq:Ess S}
\begin{vmatrix}
-\frac{5}{2\delta}+\big (c_0-\frac{17}{21}\big )\mathrm{i}k+
\frac{9\eta}{2\delta}(\mathrm{i}k)^2-\lambda &
\frac{5}{2\delta}+\frac17 \mathrm{i}
k+\frac{5}{6\delta}(\mathrm{i}k)^3-
\frac{2\eta}{\delta}(\mathrm{i}k)^2\\
& \\
-\mathrm{i}k & c_0 \mathrm{i}k-\lambda
\end{vmatrix}=0
\end{equation}
\end{widetext}
for $k\in\mathbb{R}$. The locus of the essential spectrum is shown
as a solid line in Fig.~\ref{Fig:SPC}, and coincides with the
locus of the essential spectrum of the adjoint operator (see
Appendix~\ref{AppA}), as expected. We note that part of the
essential spectrum is unstable. As it has been demonstrated in
previous studies (e.g.
Refs.~\onlinecite{Chang_JFM_1995,Tse10,Tseluiko_PD_2010}),
the unstable part of the essential spectrum is connected with the flat
film instability and can be excluded from our consideration.
\begin{figure}
\centering
\includegraphics[width=0.48\textwidth]{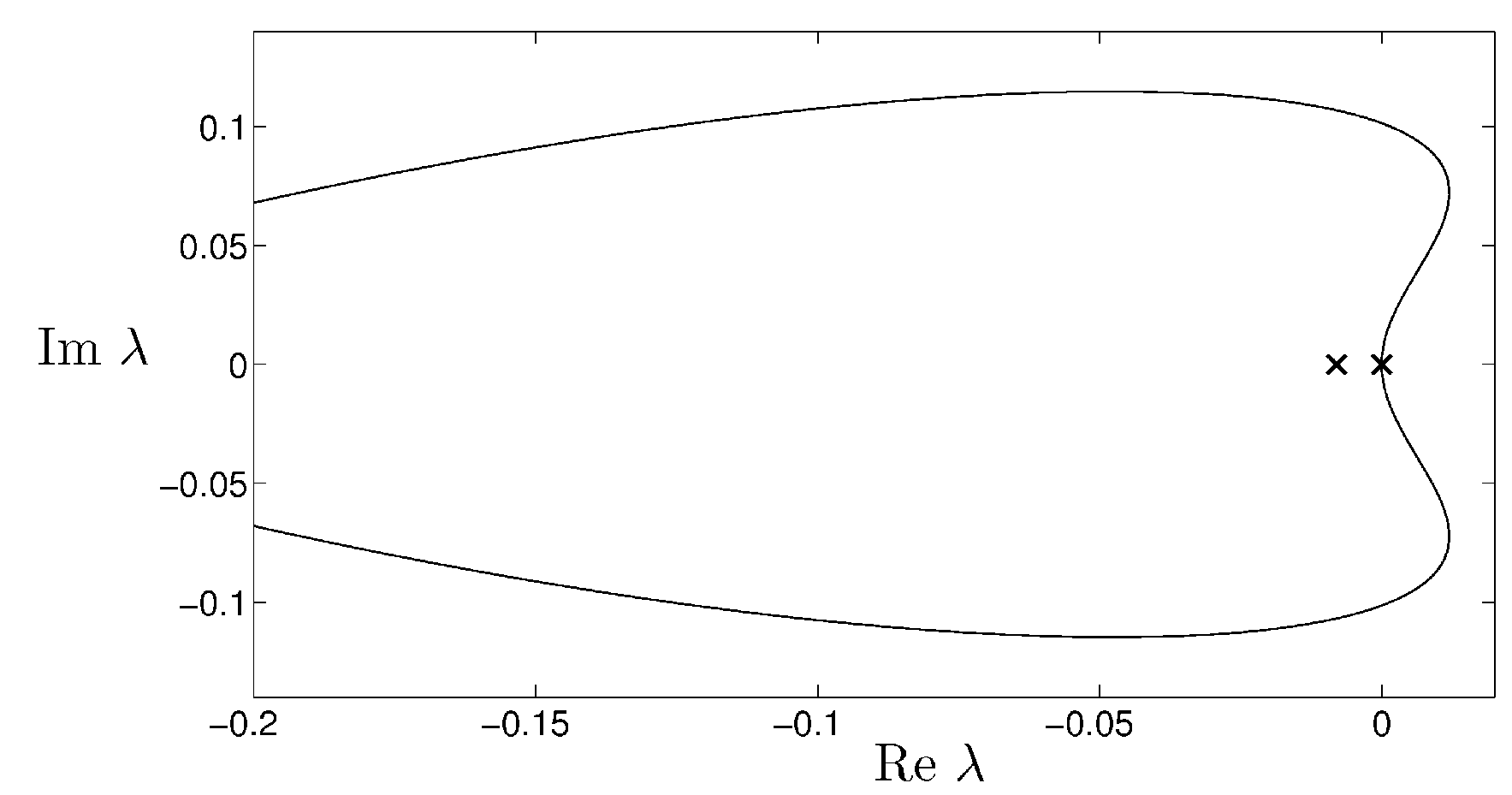}
\caption{Spectrum of $\Li$ for $\delta=0.98$ for the second-order
model with the physical parameters corresponding to W. The solid
line represents the essential spectrum and the crosses represent the
point spectrum.}\label{Fig:SPC}
\end{figure}

\begin{figure}
\centering
\includegraphics[width=0.48\textwidth]{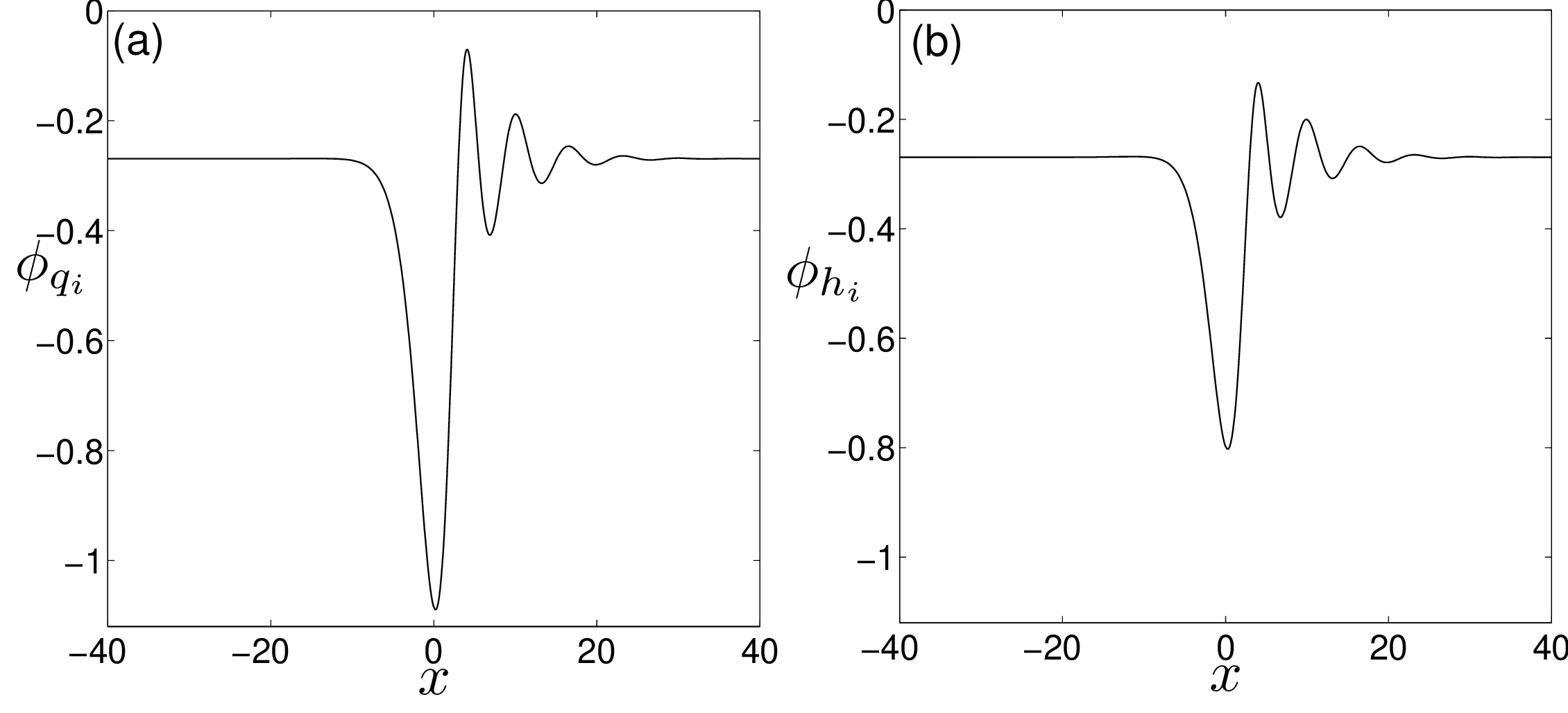}\\
\caption{Components of the generalized zero eigenfunction
$\Pht=(\phi_{q_i},\phi_{h_i})$  corresponding to the Galilean mode.
}\label{Fig:phi2}
\end{figure}

To analyze numerically the spectrum of $\Li$, we first compute the
solution for $h_i$ and $q_i$ on a finite $2L$-periodic interval by
using a pseudo-spectral method. The matrix representation of the
linear operator is obtained by applying each component of $\Li$ to a
plane wave, i.e. $\mathcal{L}_i^j \rightarrow \mathcal{L}_i^j
\mathrm{e}^{\mathrm{i}k_n x}$, for $j=1,\,\ldots,\,4$, and
transforming to Fourier space the resulting functions. By computing
this operation $\forall k_n=n\pi/L$ of the truncated Fourier series
of $h_i$, we are able to get the $n$th column of the Fourier matrix
representation of $\mathcal{L}_i^j$. We then analyze the eigenvalues
and eigenvectors for the resulting matrix.
Our results show that, in addition to the essential spectrum and the
zero eigenvalue, which is embedded into the essential spectrum,
there is one more eigenvalue, which is negative and isolated. The
eigenvalues are depicted as crosses in Fig.~\ref{Fig:SPC}. In
Fig.~\ref{Fig:phi2} we depict the two components of the
generalized eigenfunction  $\Pht$ given by the solution of
$\Li\Pht=\Pho$ and corresponds to the Galilean mode.

\begin{figure}
\centering
\includegraphics[width=0.48\textwidth]{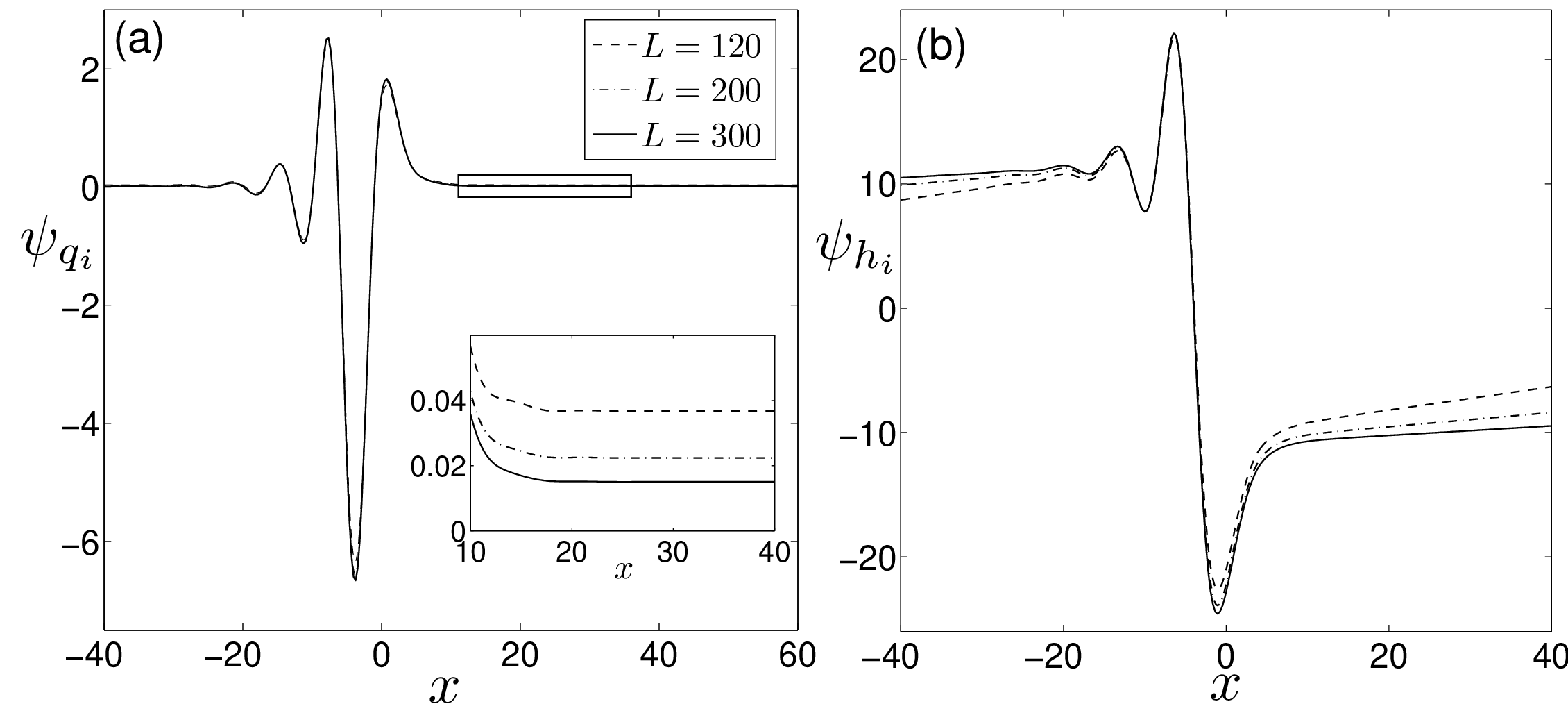}\\
\caption{Components of the generalized adjoint eigenfunction
$\Pst=(\psi_{q_i},\psi_{h_i})$ for
different values of the period $L$.}\label{Fig:psi2}
\end{figure}

On a periodic domain, the discrete part of the adjoint operator also coincides with the discrete
part of $\Li$ (see Appendix~\ref{AppA}). We find that the eigenfunction corresponding
to the zero eigenvalue on a periodic domain is merely a constant
\begin{equation}\label{Eq:Psi1}
\Pso= \binom{0}{m},
\end{equation}
and the generalized zero eigenfunction
$\Pst=(\psi_{q_i},\psi_{h_i})^T$ has to be found numerically. Its
components are shown in Fig.~\ref{Fig:psi2} for different values of
the periodicity interval.
\begin{figure}
\centering
\includegraphics[width=0.48\textwidth]{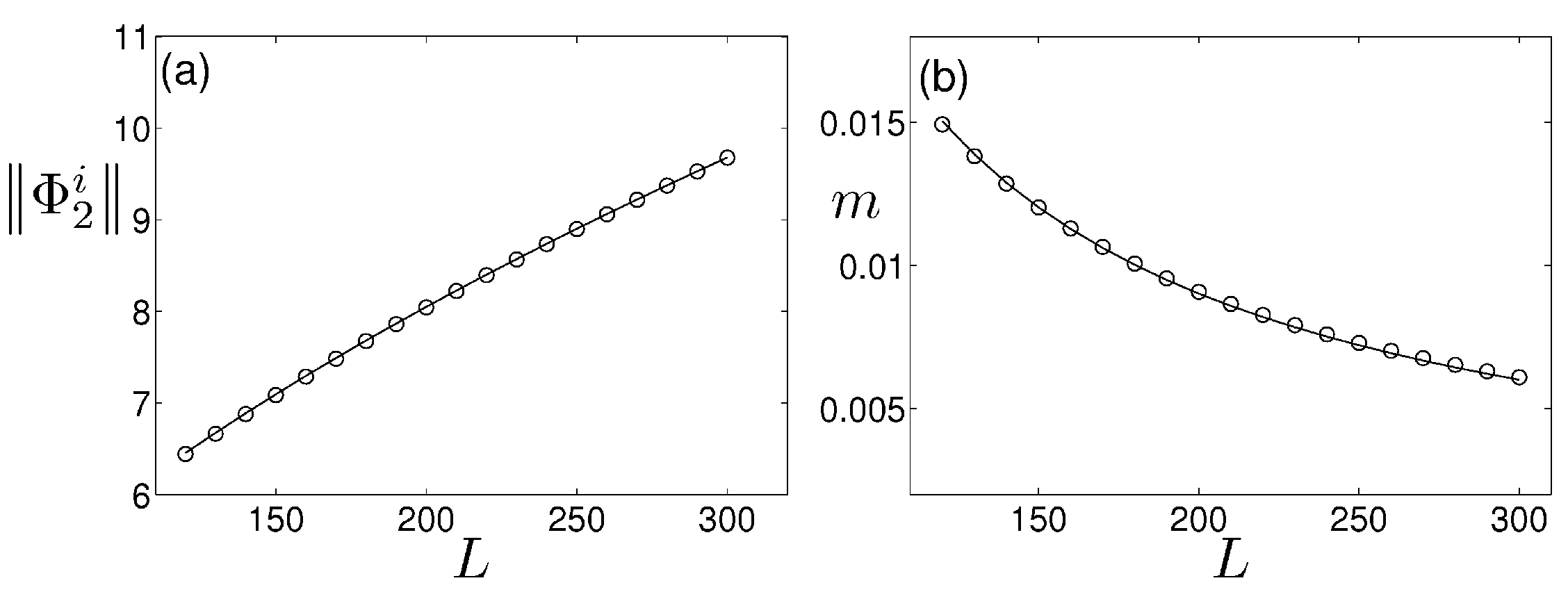}\\
\caption{(a) Norm of the generalized zero eigenfunction of $\Li$,
for different values of the system
size $L$. The solid line corresponds to a function proportional to
$L^{1/2}$. (b) Constant value $m$ of the zero eigenfunction of $\Lia$
for different $L$. The solid line corresponds to a function
proportional to $1/L$.}\label{Fig:f2}
\end{figure}
In our analysis, we have normalized the eigenfunctions so that the
following conditions hold:
\begin{eqnarray*}
& & \langle\Pho,\Pso\rangle=0, \qquad \langle\Pho,\Pst\rangle=1, {} \\
& & \langle\Pht,\Pso\rangle=1, \qquad \langle\Pht,\Pst\rangle=0,
\end{eqnarray*}
where $\langle\cdot,\cdot\rangle$ denotes the inner product
in $L_{\mathbb{C}}^{2}(-L,L)$.

We now examine the behavior of the eigenfunctions in the limit
$L\to\infty$. First, we note that both components $\phi_{q_i}$ and
$\phi_{h_i}$  corresponding to the Galilean mode $\Pht$ do not decay
to zero at infinities $x\to\pm\infty$ (cf.~Fig.~\ref{Fig:phi2}), and
therefore  we have that
$\vert\!\vert\Pht\vert\!\vert\equiv\sqrt{\langle\Pht,\Pht\rangle}\to\infty$
as the system size is increased, see Fig.~\ref{Fig:f2}(a). This
means that on an infinite domain the zero eigenvalue has both
algebraic and geometric multiplicity equal to unity and
the null space of $\Li$ is spanned only by the translational mode.
Second, we also note that the constant $m$ in (\ref{Eq:Psi1})
corresponding to  the zero eigenfunction of $\Lia$ tends to zero as
the system size is increased (see Fig.~\ref{Fig:f2}(b)), meaning
that $\Lia\Pst\to 0$ as $L\to\infty$, and thus, we have that the
function $\Pst$ belongs to the null space of $\Lia$. As it is shown
in Fig.~\ref{Fig:psi2}, the component $\psi_{q_i}$ decays to zero at
infinities for $L\to\infty$, and the component $\psi_{h_i}$ tends to
different constants as $x\to\pm\infty$, and therefore
$\vert\!\vert\Pst\vert\!\vert\to\infty$.

We then conclude that the zero eigenvalue of $\Li$ is not isolated
but belongs to the essential spectrum (and has both algebraic and
geometric multiplicity equal to unity on an infinite domain with the
null space of $\Li$ spanned by the translational mode $\Pho$). Also,
zero is not in the point spectrum of the adjoint operator, $\Lia$ on
an infinite domain (its null space is spanned by a constant and a
vector function, one component of which does not decay to zero at
infinities).
These two points complicate the
formal projections of the overlap function onto the translational
mode (the null space of $\Li$).

We remark here that we find similar spectrum features between the
second-order model and the gKS equation
\cite[][]{Duprat_PRL_2009,Tse10,Tseluiko_PD_2010}. We first note
that although the Galilean mode $\Pht$ in the gKS equation is given
by a constant, it is observed in both models that its norm tends to
infinity as $L$ is increased, and therefore it can be excluded from
the projections on the translational mode. In addition, we find the
same behaviour of the $h$-component of the generalized eigenfunction
of the adjoint operator (cf.~Fig.~\ref{Fig:psi2}$b$) and the
generalized eigenfunction of the interaction problem with the gKS
equation, namely a jump at infinity in both cases and, therefore, an
infinite norm for the corresponding eigenfunction. As was pointed
out in the coherent-structure theory for the gKS prototype
\cite[][]{Tseluiko_PD_2010}, it is possible to overcome such
difficulties by making use of a formal procedure in a weighted space
of functions that decay exponentially at $+\infty$. We shall
therefore apply such a formalism in the present study.

\subsection{Formulation in a weighted functional space} \label{subS: W space}

\begin{figure}
\centering
\includegraphics[width=0.48\textwidth]{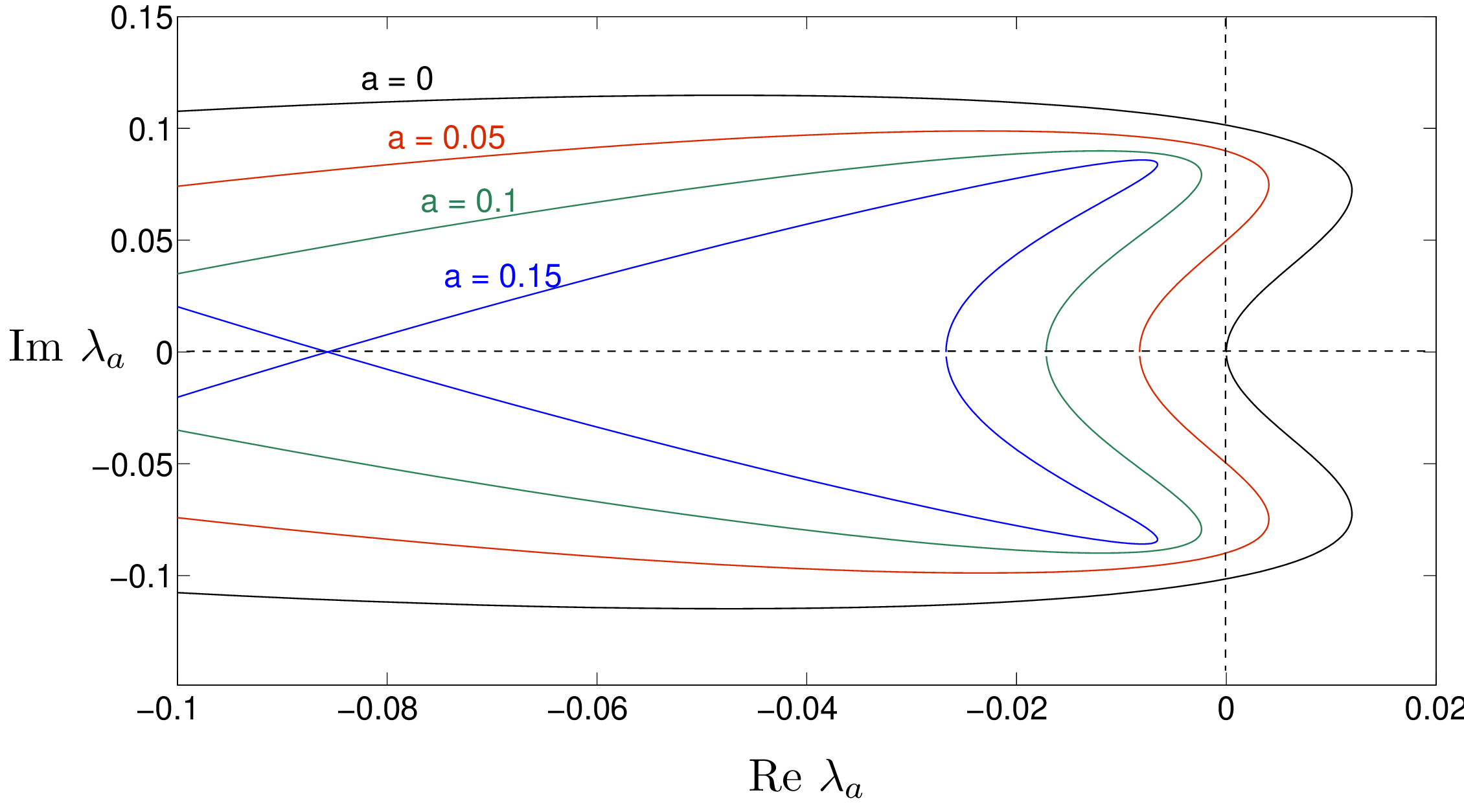}
\caption{(Color online) Essential spectrum of $\Li^a$ for $\delta=0.98$ with physical
parameters for W and different values of $a$.}\label{Fig:W SPC}
\end{figure}

Projections onto the null space can be made rigorous by choosing an
appropriate weighted space. Following the formulation used in Ref.~\onlinecite{Pego_CMP_1994}
for the Korteweg--de Vries equation and in Refs.~\onlinecite{Tse10,Tseluiko_PD_2010}
for the gKS equation, we shall
restrict our projections to the following weighted space:
\begin{equation}
 L_{a}^{2}=\{ {\boldsymbol{u}}
 :\,\mathrm{e}^{ax}{\boldsymbol{u}}\in L_{\mathbb{C}}^{2} \},
\end{equation}
with the inner product $\langle\boldsymbol{u},\boldsymbol{v}\rangle_a=
\langle\mathrm{e}^{ax}\boldsymbol{u},\mathrm{e}^{ax}\boldsymbol{v}\rangle$,
and $a>0$. The spectrum of the matrix/differential operator $\Li$ in
$L_a^2$ can be studied there through the matrix/differential
operator defined by
\begin{equation}
\Li^a\boldsymbol{u}=\mathrm{e}^{ax}\Li(\mathrm{e}^{-ax}\boldsymbol{u}),
\end{equation}
on $L_{\mathbb{C}}^{2}$. More precisely, one can easily see that the
essential spectrum $\lambda_a(k)$ of $\Li^a$ is then given by
(\ref{Eq:Ess S}) by replacing $\mathrm{i}k$ with $\mathrm{i}k-a$.
The interesting point of working in such a weighted space is that
for certain values of $a$, it is possible to completely shift the
essential spectrum to the left in the complex plane (see
Fig.~\ref{Fig:W SPC}), and therefore, the pulses may become
spectrally stable by choosing an appropriate value of $a$. Such a
shift of the essential spectrum means that the pulses are
transiently unstable but not absolutely unstable, i.e. any localized
disturbance is convected to $-\infty$ in a frame moving with the
velocity of the pulse~\cite[][]{Sandstede_Scheel_2000}. [The
essential part of the spectrum typically leads to an expanding and
growing radiation wave packet such that its left end travels
upstream and right end downstream (absolute instability). If the
speed of the right end of the wave packet is larger than that of a
solitary pulse, the pulse is destroyed, i.e. it is ``absolutely
unstable"].

It is this absolute instability of the pulses which is responsible
for the complex turbulent-like spatio-temporal chaos observed in the
KS equation, i.e. disordered dynamics in both time and space and
without clearly identifiable soliton-like coherent structures. For
the gKS equation~\cite[][]{Tse10,Tseluiko_PD_2010} there exists a
critical value of the parameter controlling dispersion (the coefficient
multiplying the third-derivative term in the equation), below which it is no longer possible to
shift the spectrum to the left half-plane and the behavior of the
gKS equation is spatio-temporal chaos like with the KS one. In our
case, however, for all $\delta$ values (or, equivalently, the
Reynolds number) we examined (up to 12) it is always possible to
shift the spectrum completely to the left half of the complex plane,
implying that dispersion effects prevent the system from evolving
into spatio-temporal chaos.

%

From the numerical point of view, having a stable essential spectrum
means that the instabilities due to numerical noise could be
eliminated and the temporal evolution of solitary waves could be
obtained for sufficiently long times. In
order to have then 
a stable essential spectrum, we substitute
\begin{equation}
 q=\mathrm{e}^{-ax}g+1/3,\qquad\qquad h=\mathrm{e}^{-ax}f+1
\end{equation}
into (\ref{Eq:Model}) to obtain the following  equations
for $g$ and $f$:
\begin{subequations}
\label{Eq:Model W space}
\begin{eqnarray}
g_t & = & c_0(g_x-ag)+\frac{5}{6\delta}f-
\frac{5}{2\delta}\frac{g-(1/3)\mathrm{e}^{-ax} f^2-(2/3)f}{h^2} \nonumber {}\\
& &-\frac{17}{7}\frac{q}{h}(g_x-ag)+\frac97\frac{q^2}{h^2}(f_x-af)
+\frac{5}{6\delta}h(f_{xxx} \nonumber {} \\
& & -3af_{xx}+3a^2f_x-a^3f)+\frac{\eta}{\delta}\bigg[4
\frac{q}{h^2}\mathrm{e}^{-ax}(f_x-af)^2 \nonumber {} \\
& &-\frac92\frac{\mathrm{e}^{-ax}}{h}(g_x-ag)(f_x-af)-6\frac{q}{h}(f_{xx} \nonumber {} \\
& & -2af_x+a^2f)+\frac92(g_{xx}-2ag_x+a^2g)\bigg],\label{Eq:Model W space a}{} \\
& & \nonumber {} \\
f_t &=& c_0(f_x-af)-(g_x-ag). \label{Eq:Model W space b}
\end{eqnarray}
\end{subequations}

Integrating numerically the above equations with an appropriate
value of $a$, ensures that the flat-film instability will not be
excited and the dynamics of the pulses will not be affected by any
numerical noise. This will be particularly useful, for instance,
when studying the interaction between two pulses, see Sec.~\ref{Sec:Two Pulses}.

The other important consequence of working in a weighted space is
that the projections onto the null space can now be made properly,
since the zero eigenvalue becomes isolated. We note that the null
space of $\Li^a$ is only spanned by the zero eigenfunction,
$\Pha=\mathrm{e}^{ax}\Pho$. Also, the zero eigenfunction of the
adjoint operator, defined  by
$\Li^{a*}=\mathrm{e}^{-ax}\Lia(\mathrm{e}^{ax}f)$, can be written as
\begin{equation} \Psa=\mathrm{e}^{-ax}(\Pst-\lim_{x\to-\infty}\Pst)
\nonumber \end{equation} which now has a finite norm. Therefore, we
will use the projection operator
\begin{equation}
 P_i(\boldsymbol{f})=\langle \boldsymbol{f},\Psa\rangle_a\Pha,
\end{equation}
for projecting onto the the null space of $\Li^a$.

\subsection{Pulse interactions and bound-state formations}

By applying the projection operator $P_i$ to (\ref{Eq:Linear})
rewritten in an exponentially weighted space and assuming that the
overlap function is ``free of translational modes" meaning that it is in the null spaces of the projections, i.e.
$P_i(\mathrm{e}^{ax}\Om)=0$, we  obtain the following dynamical
system  for the pulse locations:
\begin{equation}\label{Eq:x_dot}
\dot{x}_i(t)=
\int_{-\infty}^{\infty}\!\!\!
\big[(H_{i-1}+H_{i+1})(c_0\mathcal{J}_i^{1*}+\mathcal{J}_i^{2*})\psi_{q_i}\big]\,
\mathrm{d}x,
\end{equation}
for $i=2,\,\ldots,\,N-1$, and
\begin{eqnarray}\label{Eq:x_dot 1 N}
& &\dot{x}_1(t)=\int_{-\infty}^{\infty}\!\!\!
\big[H_{2}(c_0\mathcal{J}_1^{1*}+
\mathcal{J}_1^{2*})\psi_{q_1}\big]\,\mathrm{d}x, {} \\
& &\dot{x}_N(t)=\int_{-\infty}^{\infty}\!\!\!
\big[H_{N-1}(c_0\mathcal{J}_N^{1*}+
\mathcal{J}_N^{2*})\psi_{q_N}\big]\,\mathrm{d}x,
\end{eqnarray}
where $\mathcal{J}_i^{1*}$ and $\mathcal{J}_i^{2*}$ correspond to
the adjoint operator components of the matrix/differential operator
$\Ji$ (see Appendix~\ref{AppA}),
and $\psi_{q_i}$ is the first
component of the adjoint eigenfunction $\Pst$.
Here we have also made use of the
relation $Q_i=c_0 H_i$. If we define
the  function $\overline{\psi}_i\equiv(c_0\mathcal{J}_i^{1*}+\mathcal{J}_i^{2*})
\psi_{q_i}$, and use the notations
\begin{eqnarray}\label{Eq:S1 S2}
&& S_1(\ell)\equiv\int_{-\infty}^{\infty}\!\!H_0(x-\ell/2)
\overline{\psi}_0(x+\ell/2)\mathrm{d}x, {} \\
&& S_2(\ell)\equiv\int_{-\infty}^{\infty}\!\!H_0(x+\ell/2)
\overline{\psi}_0(x-\ell/2)\mathrm{d}x,
\end{eqnarray}
where $H_0(x)=h_0(x)-1$ and $\overline{\psi}_0$ is such that
$\overline{\psi}_i(x)=\overline{\psi}_0(x-x_i)$, (\ref{Eq:x_dot})
can be finally rewritten as
\begin{subequations}
\begin{equation}\label{Eq:x_dot S12}
\dot{x}_i(t)=S_1(x_{i+1}-x_i)+S_2(x_{i}-x_{i-1}),
\end{equation}
%
%
for $i=2,...,N-1$. For $i=1$ and $i=N$, we have
\begin{equation}
\dot{x}_1(t)=S_1(x_{2}-x_1)
\end{equation}
and
\begin{equation}
\dot{x}_N(t)=S_2(x_{N}-x_{N-1}),
\end{equation}
\end{subequations}
respectively. Therefore, the time-evolution of the $i$th pulse
location is controlled on the one hand by the function $S_1(x)$,
which describes the interaction with the monotonic tail of the
downstream pulse $i+1$, and on the other hand, by the function
$S_2(x)$ that describes the interaction with the oscillatory
capillary ripples of the upstream pulse $i-1$.

\begin{figure}
\centering
\includegraphics[width=0.38\textwidth]{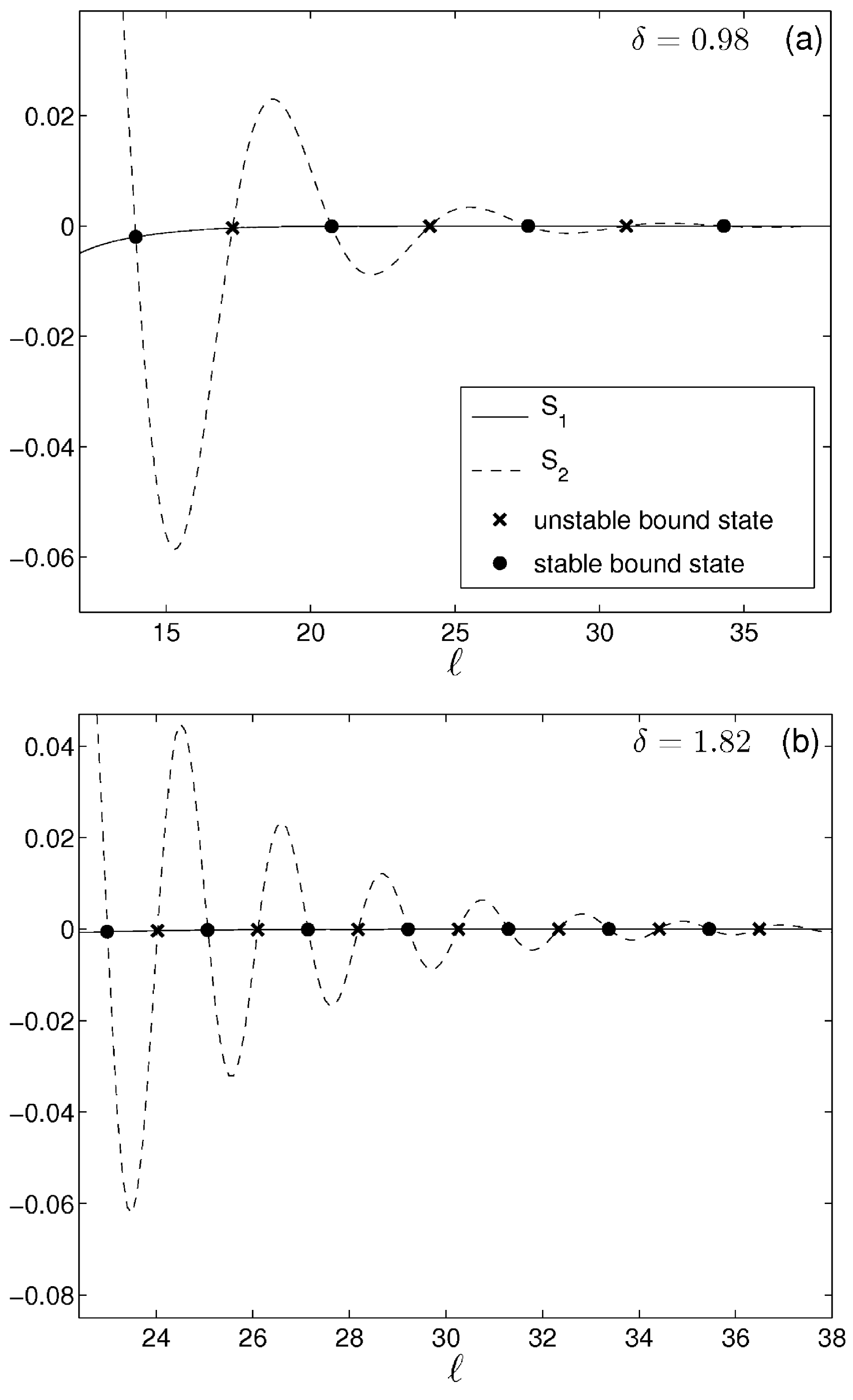}
\caption{Functions $S_1$ and $S_2$ (solid and dashed lines, respectively)
for (a) $\delta=0.98$ and (b) $\delta=1.82$. The black
circles and crosses correspond to stable and unstable bound-state separation distances,
respectively.}\label{Fig:BS theo}
\end{figure}

Let us consider the case of only two pulses interacting with each
other, i.e.~a binary interaction scenario. From Eq.~(\ref{Eq:x_dot
S12}), we can write an equation for the separation distance between
the pulses, $\ell(t)=x_2(t)-x_1(t)$, as:
\begin{equation}\label{Eq: ell theo}
 \dot{\ell}(t)=S_2(\ell)-S_1(\ell).
\end{equation}
The fixed points of the above equation are given by
\begin{equation}\label{Eq: S1=S2}
 S_1(\ell_n)=S_2(\ell_n),
\end{equation}
which predicts the different distances $\ell_n$ at which both pulses
travel at the same velocity, giving rise then to the formation of
bound states. Figure \ref{Fig:BS theo} shows the graphs of $S_1$ and
$S_2$ for two different values of $\delta$ for the second-order
model using W. Interestingly, since $S_1(\ell)$ and $S_2(\ell)$
represent the velocity of the pulses located at $x_1$ and $x_2$,
respectively, we can also predict both the velocity of the bound
state relative to $c_0$, i.e.~$c_{n}=S_1(\ell_n)$, and its
stability. Stable and unstable bound states are represented by solid
circles and crosses, respectively in Fig.~\ref{Fig:BS theo}. As long
as $S_2>S_1$, we have that $\dot{x}_2>\dot{x}_1$ so that the second
pulse moves faster than the first one leading to an increase of
$\ell$, and therefore, both pulses repel each other. On the other
hand, when $S_2<S_1$, the first pulse is moving faster than the
second one leading to a decrease of $\ell$, and therefore both
pulses attract each other.
\begin{figure*}
\centering
\includegraphics[width=0.85\textwidth]{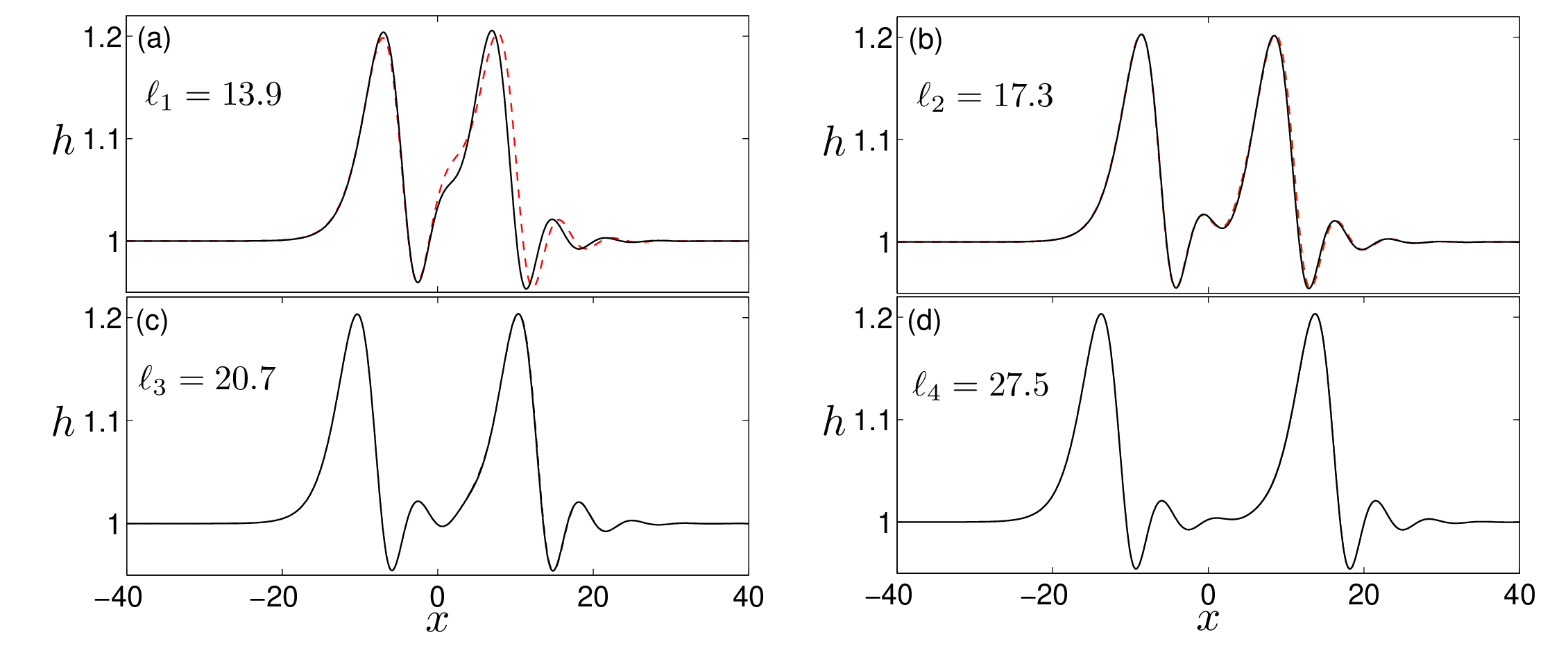}
\caption{(Color online) Numerically computed bound states (dashed lines) compared
to the theoretical predictions (solid lines) given by (\ref{Eq:theo
2 pulses}). The physical parameters correspond to W with
$\delta=0.98$.} \label{Fig:2 Pulses steady}
\end{figure*}

From a physical point of view, such oscillatory behavior of
attractions and repulsions between pulses can be understood in terms
of the interaction between the periodic capillary waves ahead of the
first pulse and the monotonic tail behind the second pulse
\cite[see][]{Duprat_PRL_2009}. More precisely, the oscillatory shape
of the free surface ahead of the hump induces a sign change of the
capillary pressure difference across the free surface, $\Delta
p_c\sim \sigma h_{xx}$. Let us consider the interaction of pulse 1
and pulse 2,  which are located at $x_1$ and $x_2$, respectively.
Note that according to our labeling, $x_1<x_2$, i.e. pulse 1 is
located to the left of pulse 2. When the tail of pulse 2 overlaps
with a maximum of one of the capillary waves of  pulse 1, there  is
a drainage process of liquid from the oscillatory tail of pulse 1 to
pulse 2 due to the positive pressure difference across the free
surface at the overlaping area between both pulses. This, in turn,
generates an increase of both the height and the speed of pulse 2,
on the one hand, and a decrease of the height and the speed of pulse
1, on the other hand. As a result, the distance between the pulses
increases corresponding to repulsion. The opposite behavior occurs
when the tail of  pulse 2 overlaps with a minimum of one of the
capillary waves of pulse 1, giving rise then to an attraction
process. When the frequency and the amplitude of the capillary
oscillations are increased by increasing $\delta$, the number of
bound states observed in a given interval  by such
attraction/repulsion mechanism increases accordingly [see
Fig.~\ref{Fig:BS theo}(b)], as expected. Note, however, that we are
always assuming that the pulses are well-separated, and although the
theory predicts bound-state formation at relatively short distances,
these are not expected to be observed.
%
\section{Numerical results}\label{Sec:num}

In this section we present extensive numerical experiments for the
second-order model (\ref{Eq:Model}). More specifically, we
investigate numerically the temporal evolution of a superposition of
two pulses as an initial condition and resulting attractive and
repulsive dynamics giving rise to formation of bound states. We
compare the numerical results with the coherent-structure theory
developed in the previous section. By imposing a localized random
initial condition, we are able to study how several pulses interact
with each other to self-organize and form bound states compound of
two or more pulses. The effect of viscous dispersion on the
bound-state formation will be elucidated by systematically
integrating both the first- and second-order models for the physical
parameters corresponding to W and WG.

\subsection{Superposition of two pulses}
\label{Sec:Two Pulses}
\begin{figure*}
\centering
\includegraphics[width=0.85\textwidth]{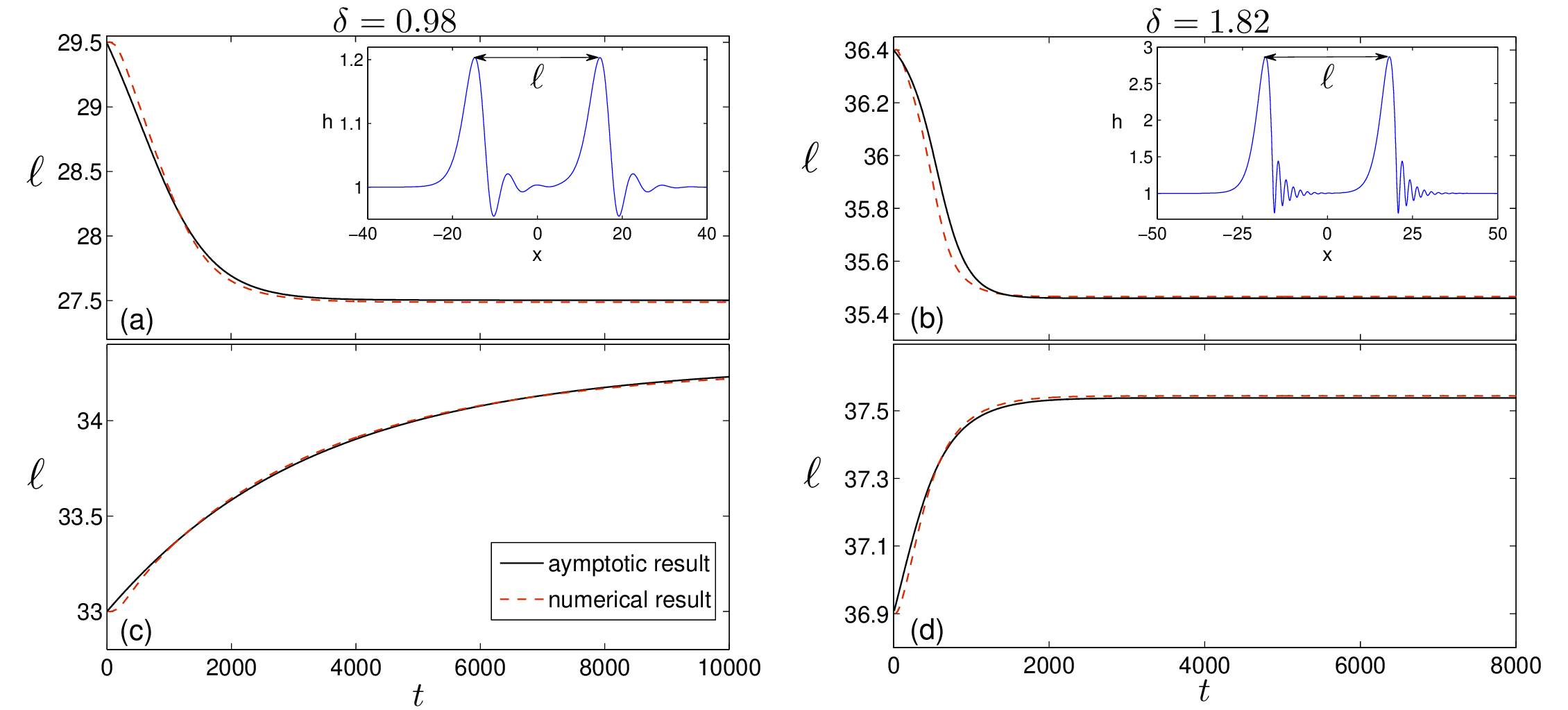}
\caption{(Color online) Evolution of the separation distance between two pulses for
 $\delta=0.98$  and $\delta=1.82$. The initial separation
distances are $\ell_0=29.5$ (a) and $33$ (c) for $\delta=0.98$, and
$\ell_0=36.4$ (b) and $33.9$ (d) for $\delta=1.82$. Solid lines
correspond to the numerical solution of the theoretical model given
by (\ref{Eq: ell theo}), and dashed lines are the numerical solution
of the second-order model in a weighted space, (\ref{Eq:Model W
space}), with $a=0.1$.}\label{Fig:2 Pulses}
\end{figure*}

Two-pulse bound states can be constructed numerically by solving
(\ref{Eq:Model Steady}) with an initial guess consisting of a
superposition of two pulses separated by the theoretically predicted
distance. We use the numerical method of Sec.~\ref{Sec:Model steady},
based on a combination of a pseudo-spectral method and Newton's
method, and the  numerical result is compared to the superposition
of two pulses,
\begin{equation}\label{Eq:theo 2 pulses}
 h=1+H_0(x+\ell_n/2)+H_0(x-\ell_n/2),
\end{equation}
where $\ell_n$ corresponds to the distances obtained by (\ref{Eq:
S1=S2}). Figure \ref{Fig:2 Pulses steady} shows the results in the
case of $\delta=0.98$ for the bound states predicted at
$\ell_1\approx 13.9$, $\ell_2\approx 17.3$, $\ell_3\approx 20.7$,
and $\ell_4\approx 27.5$ [cf. Fig.~\ref{Fig:BS theo}(a)]. As
expected, the longer separation distance between the pulses is, the
better the agreement between the theory and the numerical solution
becomes. In particular, at $\ell_3\approx 20.7$ and $\ell_4\approx
27.5$, the numerical solution and the superposition of two pulses
with the theoretically predicted bound-state separation distance are
practically indistinguishable. It is interesting to note that the
numerical scheme only converged when the pulse separation distance
for the initial guess was sufficiently close to the value predicted
by the theory. We also find good agreement between the bound-state
velocities relative to $c_0$ predicted theoretically and found
numerically: the numerical values are $c_1\approx -0.00302$,
$c_2\approx -0.00039$, and $c_3\approx-0.00006$, which are to be
compared with the theoretical predictions of $-0.00197$,
$-0.00035$, and $-0.00006$, respectively.

To check the attraction/repulsion dynamics between two pulses
predicted by the theory, we study numerically the time-evolution of
two pulses separated by an initial distance $\ell_0$ which is close
to either a stable or unstable bound-state separation distance. We
numerically integrate both the second-order model (\ref{Eq:Model W
space}) in a weighted space, and the theoretical model given by
(\ref{Eq: ell theo}). As it has been emphasized in Sec.~\ref{subS: W
space}, working in the weighted space $L_a^2$ has the advantage of
the essential spectrum being shifted to the left half of the complex
plane. Hence, any instability on the flat film region that
originates from numerical noise is eliminated, which then allows us
to follow the temporal evolution of the two pulses for sufficiently
long times. Equations (\ref{Eq:Model W space}) are integrated by
choosing $a=0.1$. To solve (\ref{Eq:Model W space}) numerically we
use the FFT to obtain the Fourier and the inverse Fourier transforms
of $g$ and $f$ in the right-hand sides of (\ref{Eq:Model W space})
and a fourth-order Runge--Kutta method to march forward in time.
A typical time step is $\Delta t=0.0025$ and the periodic domain
$[-L,L]$, with $L=120$, is discretized into 2000 intervals.

The results for $\delta=0.98$ and $1.82$ for W are presented in
Fig.~\ref{Fig:2 Pulses}. The interaction between the two pulses is
attractive when the pulses are initially separated by $\ell_0=29.5$
and $36.4$ for $\delta=0.98$ and $1.82$, respectively, converging to
the predicted stable bound states with the separation distances
approximately $27.5$ and $35.3$, respectively. Likewise, the
dynamics is repulsive when the pulses are initially separated by
$\ell_0=33$ and $36.9$ for $\delta=0.98$ and $1.82$, respectively.
The separation distances then converge to the predicted stable
bound-state separation distances $34.3$ and $37.5$, respectively. In
all cases, we found a very good agreement between the theory and the
numerical results.
\begin{figure*}
\centering
\includegraphics[width=0.85\textwidth]{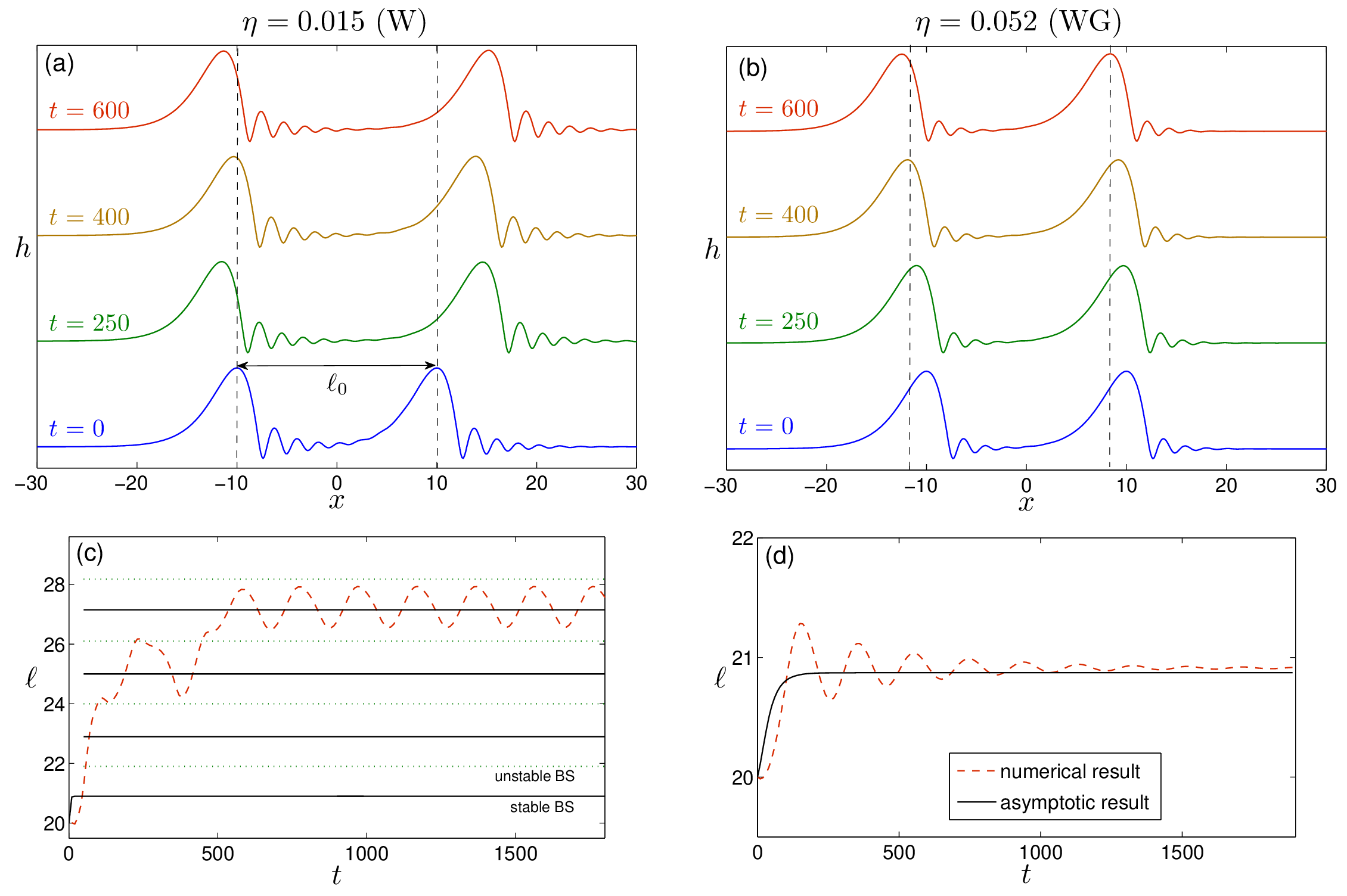}\\
\caption{(Color online) Oscillatory interaction between two pulses observed at
$\delta=1.82$ in the second-order model by using the physical
parameters of W [(a), (c)] and WG [(b), (d)]. The top panels show
the two-pulse system profile at four different times: $t=0$, $250$,
$400$, and $600$, from bottom to top, respectively. The bottom
panels show the comparison between the numerical result obtained by
integrating the second-order model in a weighted space, given by
equations (\ref{Eq:Model W space}) with $a=0.1$ (red dashed line)
and the asymptotic result predicted by the theoretical model
(\ref{Eq: ell theo}). The solid and dotted lines in (c) represent
the stable and unstable bound states (BS), respectively, predicted
by the theory. The pulses seem lock on at certain average distances
close to those predicted by the weak interaction theory.}
\label{Fig:oscillatory}
\end{figure*}

Interestingly, a different interaction behaviour emerges as both
pulses are placed close enough to each other so that the assumption
of a well-separated distance between them is not valid anymore and
the interaction is no longer weak. Figures
\ref{Fig:oscillatory}($a$) and \ref{Fig:oscillatory}($c$) depicts
the temporal evolution of two pulses separated by an initial
distance $\ell_0=20$ for a relatively high Reynolds number
($\delta=1.82$) in the second-order model using the physical
parameters for W (which gives a value of $\eta=0.015$ for the
viscous-dispersion number). We observe that both pulses attract and
repel each other in an oscillatory manner, giving rise to a rapid
fluctuating growth of the initial separation length until they reach
an oscillatory steady-state of constant frequency and amplitude [the
dashed line in Fig.~\ref{Fig:oscillatory}($c$)]. Is is remarkable
that although such an oscillatory dynamics cannot be captured by our
weak-interaction theory, the final separation length around both
pulses are oscillating corresponds to a stable bound state predicted
by the theory (solid lines), and the amplitude of the oscillations
is delimited within the distance between two consecutive unstable
bound states (dotted lines).
This type of oscillatory interaction between two pulses was first
observed by Malamataris {\emph et al.}~\cite{Malamataris_PoF_2002}
in direct numerical simulations of the full Navier--Stokes equations
with wall and free-surface boundary conditions, and was attributed
to the competition of the strong fluctuations on the capillary
pressure at the overlapping area between the pulses, and the
nonlinear response of the solitary hump when it is perturbed from
its stationary shape. In this sense, if we consider stronger viscous
dispersion effects, leading to smaller amplitude of the capillary
fluctuations in the front tail of the pulse
(cf.~Fig.~\ref{Fig:steady}), such oscillatory interaction can be
largely or completely removed. Indeed,
Fig.~\ref{Fig:oscillatory}($b$) repeats the same numerical
experiment but taking the physical parameters corresponding to WG,
which is more viscous than water and thus gives the value of
$\eta=0.052$ for the viscous dispersion number. We observe that such
oscillatory interaction is largely reduced and the two pulses
eventually reach a stable bound state in agreement with our
weak-interaction theory. This is a clear manifestation of the fact
that including the second-order viscous-dispersion effects, ignored
in all previous film-flow interaction studies, can be crucial to
obtaining an accurate description of the interaction between pulses.

\subsection{Localized random initial condition}
\begin{figure*}
\centering
\includegraphics[width=0.85\textwidth]{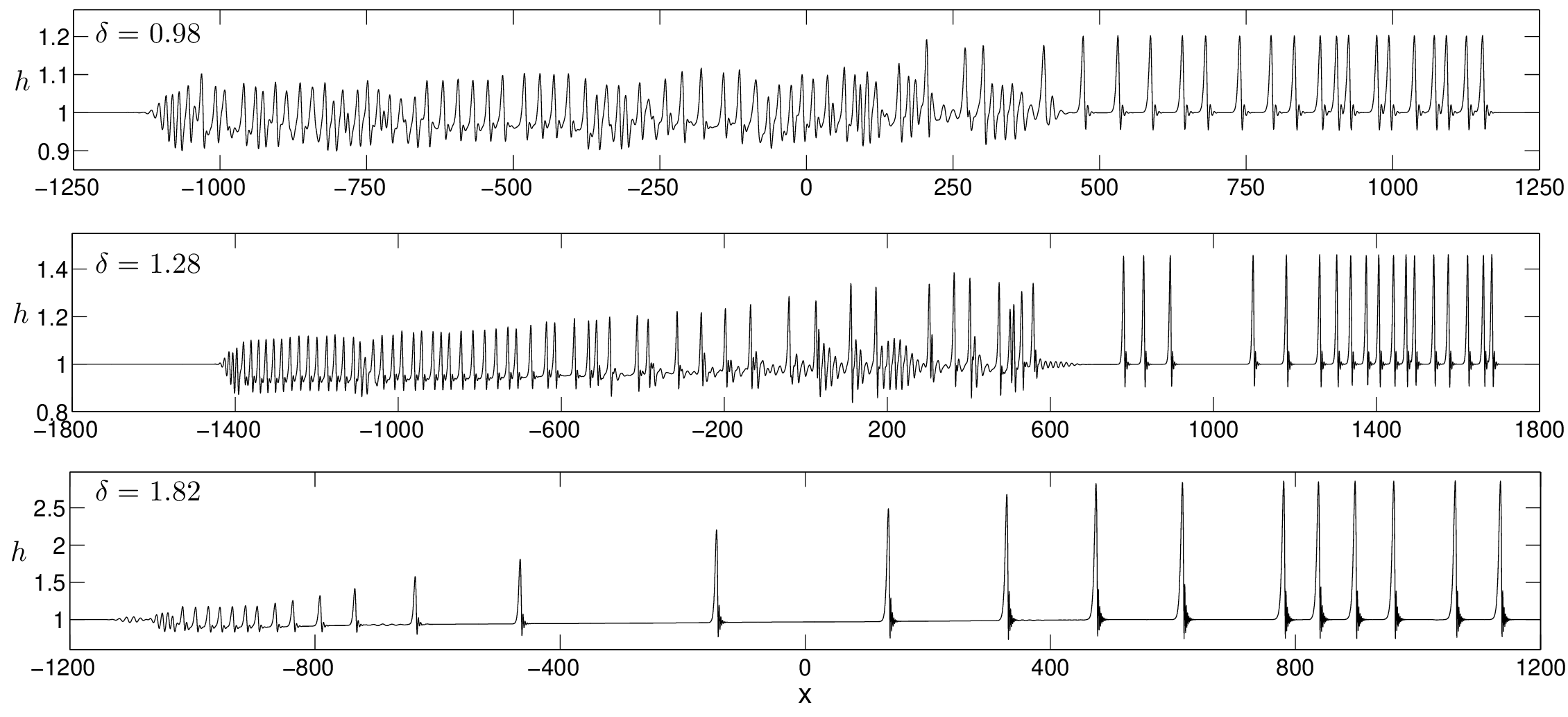}\\
\caption{Typical numerical solutions of the second-order
model~(\ref{Eq:Model}) in extended domains for $\delta=0.98,\,1.28$,
and $1.82$ at $t=2250,\,1700$, and $450$, respectively (panels (a),
(b), and (c), respectively). See supplementary material at [URL will
be inserted by AIP] for the time evolution of a localised random
initial condition for $\delta=0.98$ in the frame moving with
velocity $c_0$.}\label{Fig:unsteady}
\end{figure*}

We also integrate (\ref{Eq:Model}) by considering both the first-
($\eta=0$) and second-order model ($\eta>0$), and imposing a
localized random initial condition. In our simulations, we have used
the parameter values for both W and WG. To obtain an appropriate
random representation of a differentiable function but with a
sufficiently high frequency content we construct the initial
condition by using the following random function:
\begin{equation}
 h_\mathrm{in}(x)=\frac{\gamma_0}{\sqrt{N}}\sum_{m=1}^{N}
\bigg(\alpha_m\sin\frac{k_0 m}{N}x+\beta_m\cos\frac{k_0 m}{N}x\bigg),
\end{equation}
which has been recently proposed in Ref.~\onlinecite{Savva_PRL_2010} in the
context of wetting of disordered substrates. Here, $\gamma_0$ and
$k_0$ are the characteristic amplitude and wavenumber, respectively,
$N$ is a large positive integer, and $\alpha_m$ and $\beta_m$ are
statistically independent normal variables with zero mean and unit
variance. It can be shown that in the limit of $\gamma_0k_0\ll 1$
and $N\to\infty$, $h_\mathrm{in}(x)$ is a band-limited white noise.
To get a localized perturbation, $h_\mathrm{in}(x)$ is then
multiplied by the function $\theta(x)=[\tanh(x)-\tanh(x+L_c)]/2$,
where $L_c$ is the length of the disturbance. In our simulations we
have chosen $\gamma_0=0.05$, $k_0=1$, $N=2000$, and $L_c=500$.
\begin{figure*}
\centering
\includegraphics[width=0.85\textwidth]{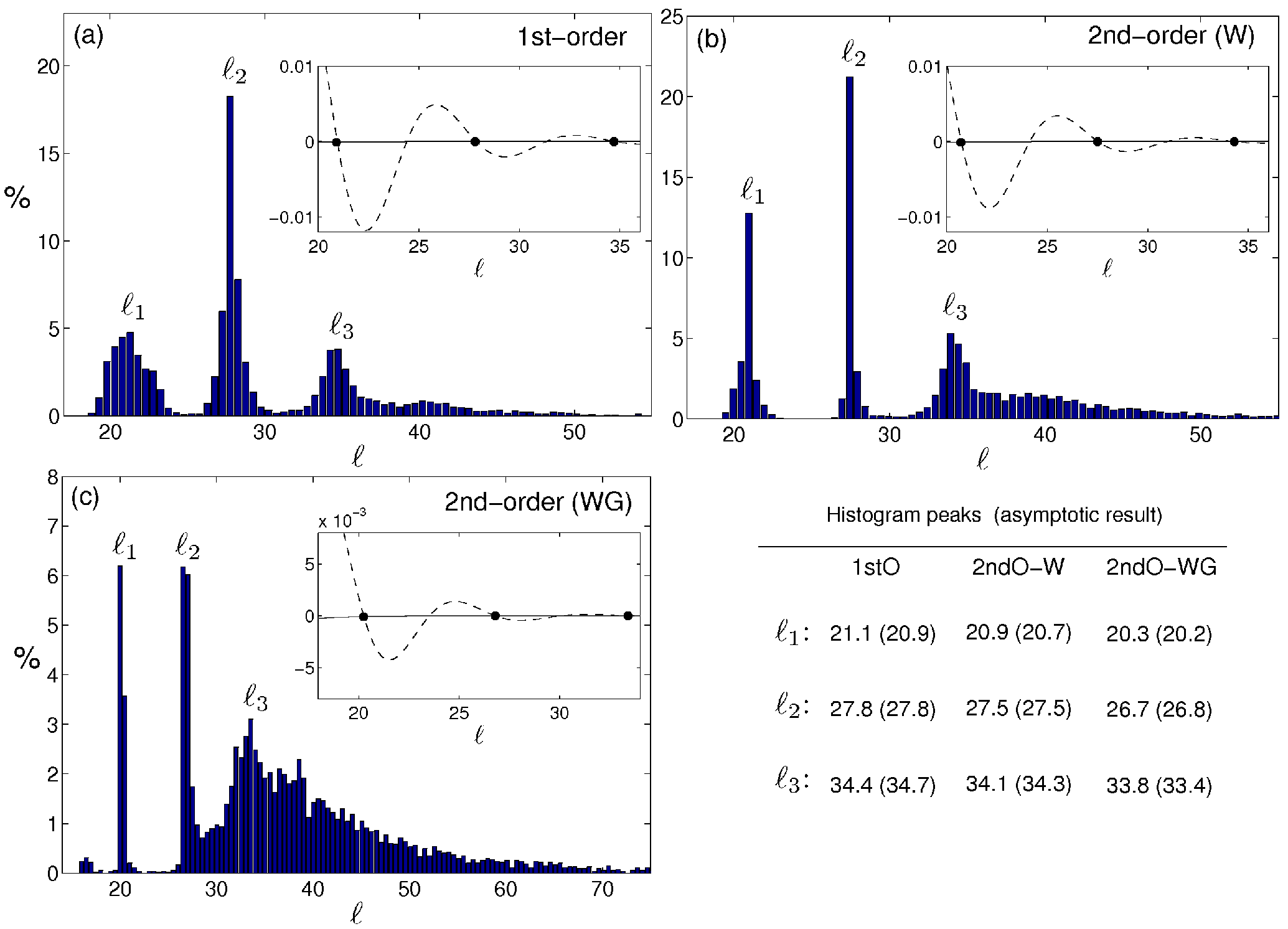}\\
\caption{(Color online) Histograms of the pulse-separation distances for $\delta=0.98$
in (a) the first-order model ($\eta=0$), (b) the second-order model with the parameter values corresponding to
W ($\eta=0.011$), and (c) the second-order model with the parameter values corresponding to WG ($\eta=0.041$).
The table shows the locations of the peaks observed in the
histograms obtained for the 1st-order (1stO) and 2nd-order (2ndO) models
compared to the values predicted by (\ref{Eq:x_dot S12}),
plotted in the insets of each panel, where the dashed and solid lines
correspond to the $S_2$ and $S_1$ functions, respectively.}
\label{Fig:Hist del1}
\end{figure*}
\begin{figure*}
\centering
\includegraphics[width=0.85\textwidth]{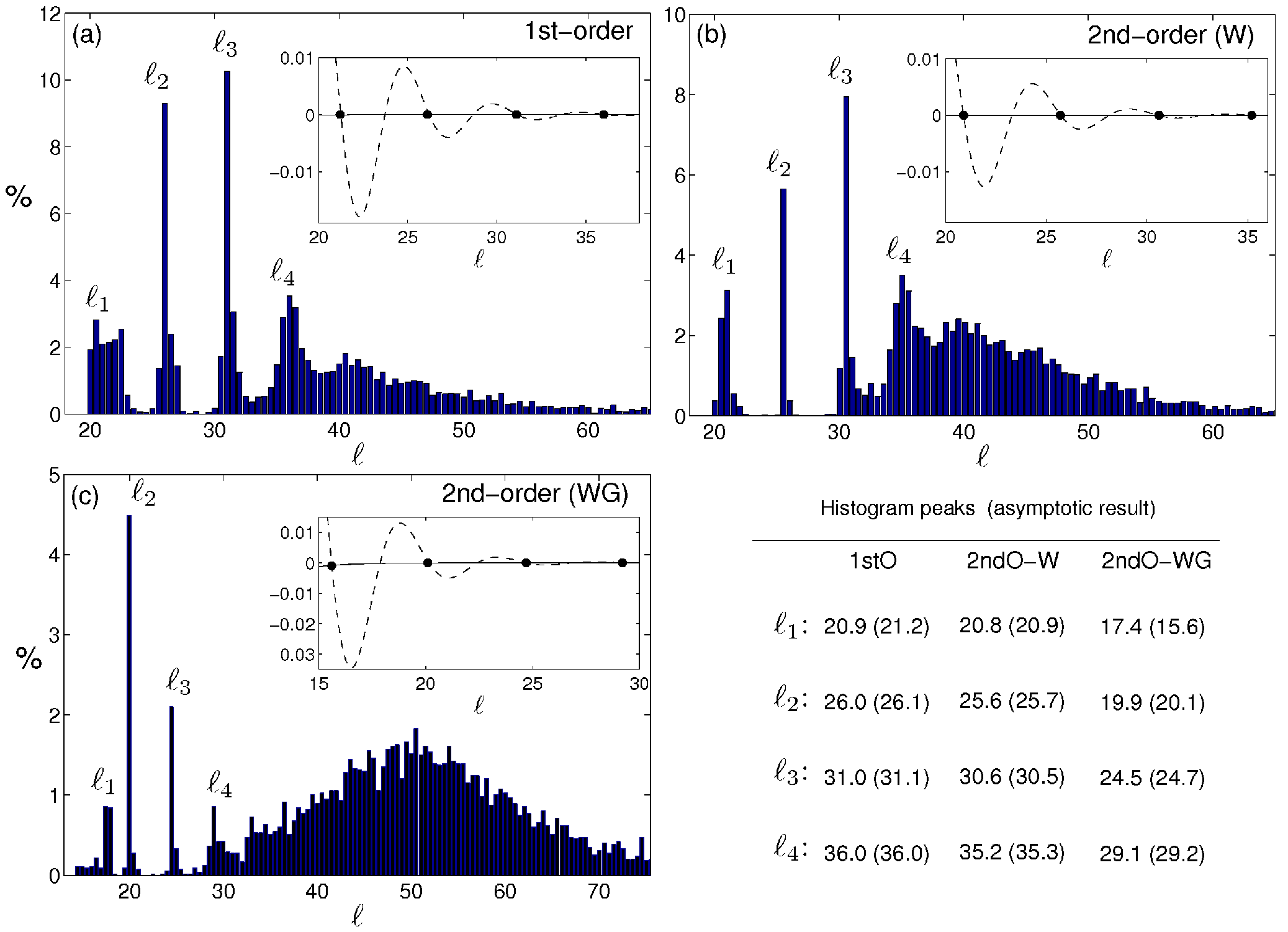}\\
\caption{(Color online) Histograms of the pulse-separation distances for
$\delta=1.28$ in (a) the first-order model ($\eta=0$), (b) the
second-order model with the parameter values corresponding to W
($\eta=0.013$), and (c) the second-order model with the parameter
values corresponding to WG ($\eta=0.046$). The table shows the
locations of the peaks observed in the histograms compared to the
values predicted by the model given by (\ref{Eq:x_dot S12}), plotted
in the insets of each panel, where the dashed and solid lines
corresponds to the $S_2$ and $S_1$ functions, respectively.}
\label{Fig:Hist del2}
\end{figure*}

To accelerate the computational efficiency of our numerical scheme,
we have used a Runge--Kutta--Fehlberg method with dynamic time-step
adjustment. At each time step this scheme computes two different
solutions and compares them. The time-step is the redefined
according to the agreement between both solutions. We impose a
minimum time step $\Delta t=0.001$ (that is reached only in
computations for large values of $\delta$), and we are able to use
time steps as large as $\Delta t=0.01$. We integrated equations
(\ref{Eq:Model}) on a periodic domain $[-L,L]$ discretized into $2M$
intervals by using the FFT to obtain the Fourier and the inverse
Fourier transforms of $q$ and $h$ in the right-hand sides of (\ref{Eq:Model}).
Note that the use of periodic boundary conditions turns out to be quite
convenient to solve unforced systems in the frame moving with
velocity $c_0$.
We used the following values: $L=1250$, $1800$, and
$1200$ and $M=2000$, $3000$, and $4000$ for $\delta=0.98$, $1.28$,
and $1.82$, respectively. The initial disturbance results in an
expanding wave packet whose left envelope travels upstream (absolute
instability; see our comment in Sec.~\ref{subS: W space}) and the
right one travels
with a velocity lower to that of an individual
pulse.
The wave packet grows as it propagates and gives
birth to a number of pulses escaping the expanding wave packet. The
equations were integrated up to $t=2250$, $1700$, and $450$ for
$\delta=0.98$, $1.28$, and $1.82$, respectively. Beyond these times
the rear side of the expanding packet starts to interact with the
front pulses. Figure \ref{Fig:unsteady} shows typical solutions for
the different values of $\delta$  by using the physical parameters
of W in the second-order model. In order to compute the histograms of
the pulse-separation distances for $\delta=0.98$ and $1.28$, we took into
account the first 12 pulses which are located at the righmost part of the
corresponding panels of Fig.~\ref{Fig:unsteady}, and performed 600 different
realisations, giving a total number of 6600 separation lengths. For $\delta=1.82$, we
took into account the first 6 pulses, and performed 1200 realisations, giving
6000 separation lengths in total.

Figure~\ref{Fig:Hist del1} depicts the results obtained for the
first- and second-order models at $\delta=0.98$ for W and WG at
$\Rey=3$ and $1.59$, respectively. Panel (a) shows the histogram of
the separation distances between pulses obtained from the
first-order model and panels (b) and (c) from the second-order model
for W and WG, respectively. The inset in each panel shows the
corresponding $S_1$ and $S_2$ functions from the theoretical model
given by (\ref{Eq: ell theo}).
We first note that in all the cases the distributions for the pulse
separation distances are mainly characterized by three peaks that
correspond in each case to the theoretically predicted distances at
which two-pulses bound states are formed (see table in
Fig.~\ref{Fig:Hist del1}). \emph{It is interesting to note that the 
peaks appear to be broad, indicating that dynamic interaction between 
pulses seems to persist indefinitely. Such interaction, however, is 
affected by viscous dispersion, obtaining that the peaks of the 
distributions become more pronounced and sharper for the simulations 
of the second-order model, specially the
ones observed at short distances, $\ell_1$ and $\ell_2$ [cf.~\ref{Fig:Hist del1}(b) 
and \ref{Fig:Hist del1}(c)]}. Also, while
the locations at which bound-states are formed
are approximately  the same in all cases, the dominant peaks in the
distribution change as the viscous-dispersion effects are increased.
In particular, we observe that the distribution for the first-order
simulations is mainly dominated by the bound states formed at
$\ell_2$, whereas the distributions for the second-order simulations
are dominated by the peaks at $\ell_1$ and $\ell_2$, particularly in
the WG case where both peaks have approximately the same
probabilities. We also note a small peak, less than $0.5\%$, in the
second-order WG results, reflecting that the shortest
pulse-separation distance can be affected by  second-order terms.
A more clear picture on the effect of viscous dispersion on pulse dynamics
is expected to emerge as $\delta$ is increased.
\begin{figure*}
\centering
\includegraphics[width=0.85\textwidth]{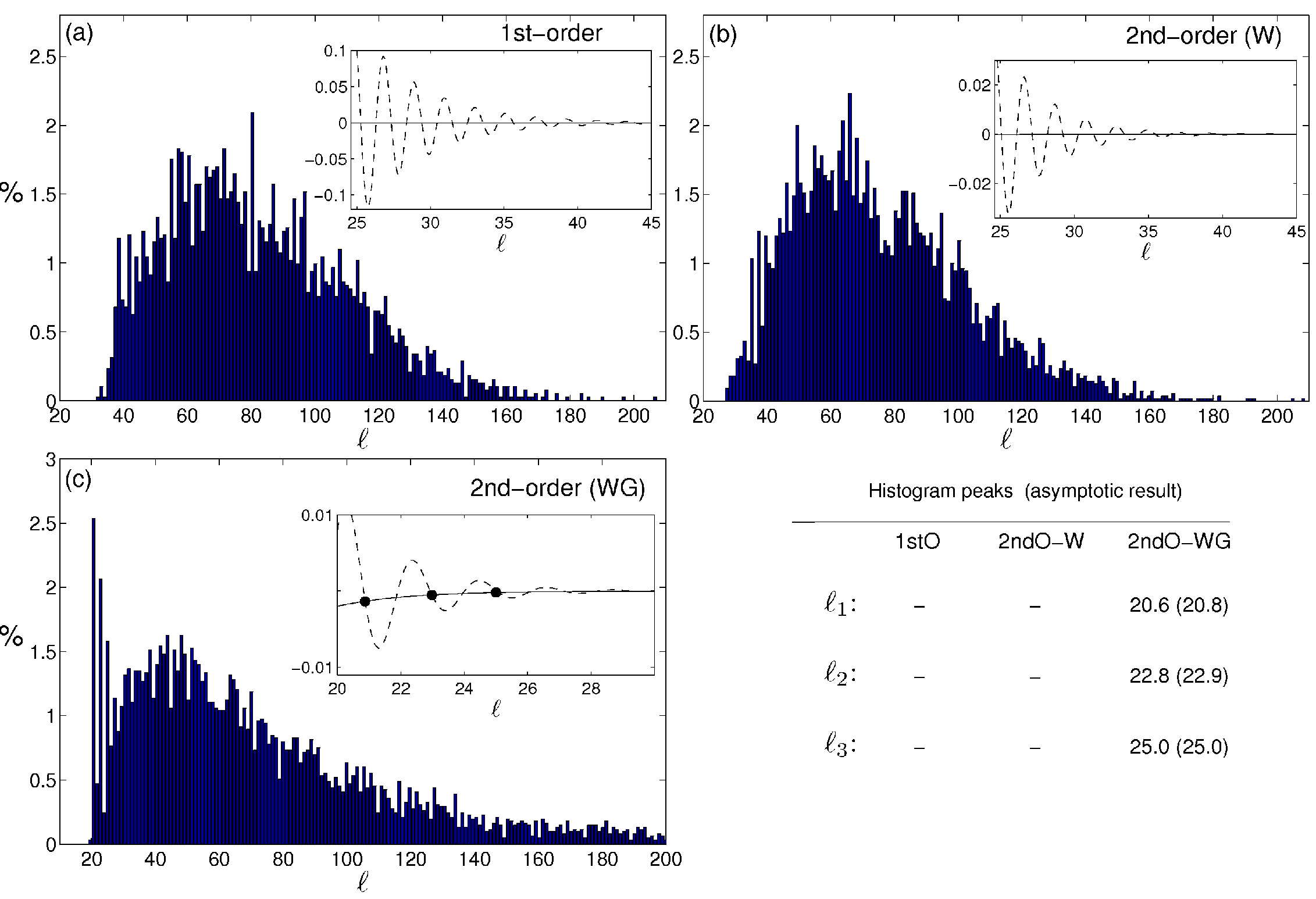}\\
\caption{(Color online) Histograms of the pulse-separation distances for
$\delta=1.82$ in (a) the first-order model ($\eta=0$), (b) the
second-order model with the parameter values corresponding to W
($\eta=0.015$), and (c) the second-order model with the parameter
values corresponding to WG ($\eta=0.052$). The table shows the
numerical value of the peaks observed in the histograms compared to
the values predicted by the model given by (\ref{Eq:x_dot S12}),
plotted in the insets of each panel, where the dashed and solid
lines corresponds to the $S_2$ and $S_1$ functions, respectively.}
\label{Fig:Hist del3}
\end{figure*}

Figure \ref{Fig:Hist del2} shows the results for $\delta=1.28$, that
corresponds to $\Rey=3.75$ and $2$ for W and WG, respectively. The
distributions are again characterized by the presence of peaks
representing the selected distances at which bound states are
observed. In this case, we observe four peaks that are sharper and
more pronounced in the second-order simulations than we observed
before. We find that the histogram peaks obtained in the
W-simulations occur at similar distances for both the first- and the
second-order models, and are in very good agreement with the theory
(see the table in Fig.~\ref{Fig:Hist del2}). It is interesting to
note, however, that such a similarity between both models is no
longer observed when the simulations are made by using the parameter
values for WG. In this case, the second-order effects start to
become important, and the distances at which the bound states are
observed change accordingly. We find that the first observed peak
for the second-order model occurs at $\ell_1\simeq 17.4$, in
contrast to the value obtained by using the first-order model,
$\ell_1\simeq 20.9$. Also, we note that the dominant peak in the
histogram for the second-order model is found at $\ell_2\simeq
19.9$, whereas the peak corresponding to the first-order model is
found at $\ell_3\simeq 31$. All of these differences can be
explained in terms of the decrease of the amplitude of the capillary
ripples preceding the main solitary humps that is largely influence
by second-order effects (cf. Fig.~\ref{Fig:steady}), which in turn,
allows the pulses to get closer to each other. It is worth
mentioning that for $\ell_1\approx 17.4$ the assumption of
well-separated pulses is not quite satisfied and, as expected, the
theoretical and numerical results are not in good agreement.

Figure~\ref{Fig:Hist del3} depicts the results for $\delta=1.82$,
that corresponds to $\Rey=5$ and $2.66$ for W- and WG-simulations,
respectively. The W-simulations reveal that there is no any clear
distance selection among pulses. The distributions of the separation
distances for both the first- and second-order models are broader,
as compared to the previous cases and without a dominant peak. This
seems to indicate that the pulses start to feel the effect of some
underlying chaotic behaviour with the eventual loss of
self-organisation into bound states, but unlike the spatio-temporal
chaos with the KS equation, solitary pulses are still clearly
identifiable.
Interestingly, when the simulations are performed using the values
of the parameters corresponding to WG, the consequence of including
the viscous-dispersion effects due to the second-order terms becomes
crucial in the bound-state formation process. Indeed, the histogram
obtained for the second-order model is still dominated by the
presence of three peaks which are in agreement with the
theoretically predicted bound states. This is a clear evidence that
viscous dispersion plays an important role in the self-organisation
process that brings the system in a state which can be described in
terms of bound states.

\subsection{Solitary-wave gas density}

Our numerical observations for $\delta=0.98$ and $1.28$ have shown
that there is a clear selection process for which the pulses
self-organise to form bound states described by well-defined
distances. From a statistical point of view, it is therefore
reasonable that in a large domain containing many solitary waves
interacting through attractions and repulsions with each other, 
the system can be described in terms of
a mean separation distance, that gives information about  the
density of solitary waves.
In this sense, we can treat the system as
a ``gas" compound of solitary waves with a well-defined density.
To characterize such a \emph{solitary-wave gas}, we shall assume that
for long times and large spatial domains, the distribution of the
separation distances between pulses is mainly dominated by the peaks
observed in Figs.~\ref{Fig:Hist del1} and \ref{Fig:Hist del2}.
This is a reasonable assumption considering that the peaks become more
pronounced as the system evolves on time, something that has been observed
in the gKS equation~\cite{Tse10,Tseluiko_PD_2010}, and also in our numerical
experiments. We then define the \emph{mean separation distance} as
\begin{equation}
 \langle\ell\rangle= \sum_{i=1}^{n}\alpha_i\ell_i,
\end{equation}
where $n$ is the number of peaks of the histogram, and $\alpha_i$ is
an average weight that we approximate as
\begin{equation}
 \alpha_i=\frac{p_i}{\sum_{1}^{n}p_i},
\end{equation}
where $p_i$ is the probability of each peak. By using this
definition we can estimate the typical mean separation length
between the pulses and therefore, we can obtain an estimate of the
density of the solitary-wave gas by using:
\begin{equation}
\rho_s=\frac{1}{\langle\ell\rangle}.
\end{equation}
\begin{table}
  \begin{center}
  \begin{tabular}{l|cc}
                & $\delta=0.98$   &   $\delta=1.28$ \\
      \hline
       1st-order     & 0.036           &   0.034 \\
       2nd-order (W)   & 0.038           &   0.035 \\
       2nd-order (WG)  & 0.039           &   0.045 \\
  \end{tabular}
\caption{Density $\rho_s$ of the solitary-wave gas for different
values of $\delta$ and for the first-order model, and
second-order model for both W and WG.}
  \label{tab:density}
  \end{center}
\end{table}
The densities obtained for each $\delta$ are given in Table
\ref{tab:density}. As expected, the density of the solitary-wave gas
increases as the viscous-dispersion effects are taken into account. 
It is important to emphasize that such a density has been defined
from a selected preferential mean separation length, and not from a
random mean length that would arise, for instance, if the pulses
were irregularly spaced. This is an important point which reflects
how a quasi-turbulent system (in the sense that the pulses are
continuosly and randomly interacting with each other) has an underlying
ordered, or a ``permanent`` self-organized state, that can be
understood in terms of bound-state formation.
An important feature of this state, the average separation distance
between the pulses, is largely dependent on viscous-dispersion
effects.

\section{Conclusions}\label{Sec:concl}

We have examined both analytically and numerically the interactions
of two-dimensional solitary pulses in falling liquid films. We
focused in particular on the formation of bound states of pulses and
how it is affected by the second-order (in the long-wave expansion
parameter) viscous-dispersion effects. To this end, we made use of a
second-order two-field model derived in
Refs.~\onlinecite{RuyerQuil_EPJB_2000,Ruy02}. This is a system of coupled
nonlinear partial differential equations for the local flow rate and
the film thickness.

Our theoretical investigation of the formation of bound states was
based on appropriately extending the rigorous coherent-structure
theory for the gKS equation recently developed in
Refs.~\onlinecite{Duprat_PRL_2009,Tse10,Tseluiko_PD_2010} to the
second-order two-field model (and thus putting the
coherent-structure theory for falling films on a rigorous basis). By
assuming that the solution is given by a superposition of $N$ pulses
and a small overlap (correction) function, i.e. assuming that the
pulses are well separated (weak interaction), we were able to write
down a dynamic equation for the overlap function in the vicinity of
each pulse, that is described by a linear matrix/differential
operator and contains forcing terms due to the neighboring pulses. A
careful and detailed analysis of the spectral properties of the
linear operator revealed that the relevant eigenfunction is the
translational mode that corresponds to the zero eigenvalue. This
eigenvalue is not isolated and, therefore, belongs to the essential
spectrum. The null space of the adjoint operator is spanned by a
constant vector function and another non-constant function that has
an infinite norm, meaning that zero is not in the point spectrum of
the adjoint operator. This spectral behavior is similar to the one
observed for the gKS equation \cite[][]{Tse10,Tseluiko_PD_2010}.
Projections to the translational mode were made rigorous by using
formulation in a weighted space. The outcome of the projections is a
dynamical system for the pulse locations. By studying its fixed
points, we were able to predict the distances at which the bound
states are formed.

Numerical experiments of the temporal evolution of a superposition
of two pulses have been found in very good agreement with the
theoretical predictions, and in particular, the theoretically
predicted attractive and repulsive dynamics that gives rise to the
formation of bound states. We demonstrated that the second-order
viscous-dispersion terms are crucial for an accurate description of
the pulse interactions. These terms affect the amplitude and
frequency of the capillary ripples in front of a solitary pulse. So
their influence is in fact linear, but interestingly they can have
some crucial consequences on the nonlinear dynamics of the film and
the wave-selection process in the spatio-temporal evolution. After
all, solitary pulses interact through their tails which overlap,
i.e. the capillary ripples and their amplitude and frequency will
affect the separation distance between the pulses: for example,
smaller-amplitude ripples will allow for more overlap between the
tails of neighboring pulses, thus decreasing their separation
distance. This in turn will affect the average separation distance
between pulses when the system reaches its permanent quasi-turbulent
regime and hence the density of the solitary waves. This also means
that any model that does not include second-order terms should be
used with caution and certainly not for an accurate description of
solitary pulse interaction which dominates the spatio-temporal
dynamics of the film.

In addition, we have studied strong interaction between two pulses
and found that it leads to an oscillatory behavior of the separation
length as was also numerically observed in
Ref.~\onlinecite{Malamataris_PoF_2002} from direct numerical
simulations of full Navier--Stokes equations and wall and
free-surface boundary conditions. This behavior escapes the
description of our weak-interaction theory. Again, we find that
viscous dispersion effects are crucial and the strong nonlinear
interaction between pulses can be largely reduced by increasing the
viscous-dispersion parameter.

We have also performed numerical simulations in extended domains
with a localized random initial condition. These allowed us to study
the interaction between the pulses and how they self-organize to
form bound states in extended domains. Detailed statistical analysis
of the pulse separation distances revealed that the histograms of
the separation distances have clear peaks that correspond to the
theoretically predicted bound states. We have used different values
of the reduced Reynolds number $\delta$ and two different sets of
physical parameters corresponding to two liquids of different
viscosities, namely water and a mixture of water and glycerin. In
all cases, we have observed that the peaks of the histograms
corresponding to the shortest distances are always more pronounced
in the second-order simulations. As expected, the differences
between the first- and the second-order models were found to be more
important in the case of the higher viscosity liquid. We observed
that the minimum distance at which bound states are formed is always
shorter in the second-order simulations. It is important to note
that in the cases of $\delta=0.98$ and $1.28$, both models have
shown that there is always formation of bound states, indicating
that statistically, the free surface can be treated as a
solitary-wave gas, characterized by a typical constant mean distance
between the pulses.

Of particular interest would be the extension of the coherent
structures theory developed here to other viscous flow problems, for
example the problem of a thin film coating a vertical
fiber~\cite[e.g][]{Kal94,Dup07,Ruy08} in which case due to the
Rayleigh--Plateau instability the film breaks up into a train of
droplike solitary waves. The recent experiments in
Ref.~\onlinecite{Ruy08} indicate clearly the formation of bound
states in a region of the parameter space where the
Rayleigh--Plateau instability competes with viscous dispersion. The
qualitative agreement between the experiments and the
coherent-structure theory developed in
Refs.~\onlinecite{Tse10,Tseluiko_PD_2010} is encouraging. Our hope
is that quantitative agreement can be achieved by extending the
present theory to the second-order model for flow down a fiber
developed in Ref.~\onlinecite{Ruy08}. For that matter, the present
formalism could be extended to other two-equation systems, e.g.
coupled KS-type equations used to describe synchronization
phenomena~\cite{Tasev2000}.

Finally, as noted in the Introduction, the two-dimensional pulses
considered here are only observed up to a certain, low-to-moderate,
value of the Reynolds number. At higher Reynolds numbers,
two-dimensional pulses develop an instability in the transverse
direction and a transition to a fully developed three-dimensional
regime is observed (e.g.~\onlinecite{Dem07,Dem07b}). The
coherent-structure theory developed here can be viewed as a
foundation first step for the analysis of the interactions between
three-dimensional pulses in falling films.

\begin{acknowledgments}
We thank Christian Ruyer-Quil for stimulating discussions on falling
films, Nikos Savva for useful discussions concerning the random
initial condition of the time-dependent computations and Vasilis
Bontozoglou for suggesting to us the possibility of oscillatory
interaction between two pulses. We acknowledge financial support
from EU-FP7 ITN Multiflow and ERC Advanced Grant No. 247031.
\end{acknowledgments}

\appendix
\section{The adjoint operators}\label{AppA}

The adjoint operator $\Lia$ is defined as
\begin{equation}
\langle \boldsymbol{u},\Li\boldsymbol{v}\rangle=
\langle \Lia\boldsymbol{u},\boldsymbol{v}\rangle,
\end{equation}
where $\langle \cdot,\cdot\rangle$ denotes the usual inner product
in $L_{\mathbb{C}}^{2}$ given by
\begin{equation}
\langle\boldsymbol{u},\boldsymbol{v}\rangle=\int_{-\infty}^{\infty}
\!\!\!\!\!\boldsymbol{u}^{T}\!\cdot\overline{\boldsymbol{v}}\, \mathrm{d}x.
\end{equation}
After integration by parts, we find that the adjoint operator is
\begin{equation}
 \Lia=\begin{pmatrix}
  \mathcal{L}_i^{*1} & \mathcal{L}_i^{*2} \\
   \mathcal{L}_i^{*3}& \mathcal{L}_i^{*4}
\end{pmatrix},
\end{equation}
with components
\begin{eqnarray*}
\mathcal{L}_i^{*1} & = & -\frac{5}{2\delta}\frac{1}{h_i^2}+\frac{1}{7}
\frac{q_i}{h_i^2}h_{ix}-\bigg(c_0-
\frac{17}{7}\frac{q_i}{h_i}\bigg)\partial_x\nonumber {} \\
& & +\frac{\eta}{\delta}\bigg(-\frac32\frac{h_{ix}^2}{h_i^2}-\frac12
\frac{h_{ixx}}{h_i}+\frac92\frac{h_{ix}}{h_i}\partial_x+
\frac92\partial_{xx}\bigg),  {} \\
\mathcal{L}_i^{*2}& = & \partial_x {} \\
\mathcal{L}_i^{*3} & = &  \frac{5}{6\delta}+\frac{5}{\delta}\frac{q_i}{h_i^3}-
\frac{1}{7}q_i\frac{q_{ix}}{h_i^2}+ \frac97
\frac{q_i^2}{h_i^2}\partial_x -\frac{5}{2\delta}h_{ix}\partial_{xx} {} \\
&&-\frac{5}{6\delta}h_i\partial_{xxx}\nonumber + \frac{\eta}{\delta}\bigg[4 q_{ix}\frac{h_{ix}}{h_i^2}-
4q_i\frac{h_{ix}^2}{h_i^3}+4q_i\frac{h_{ixx}}{h_i^2} {} \\
&&-\frac32\frac{q_{ixx}}{h_i}-\bigg(20q_i
\frac{h_{ix}}{h_i^2}+\frac{15}{2}\frac{q_{ix}}{h_i}\bigg)\partial_x-
6\frac{q_i}{h_i}\partial_{xx} \bigg], {} \\
\mathcal{L}_i^{*4} & = & -c_0\partial_x.
\end{eqnarray*}
The spectrum of the adjoint operator is shown in Fig.~\ref{Fig:SPC
A}.
%
\begin{figure}
\centering
\includegraphics[width=0.48\textwidth]{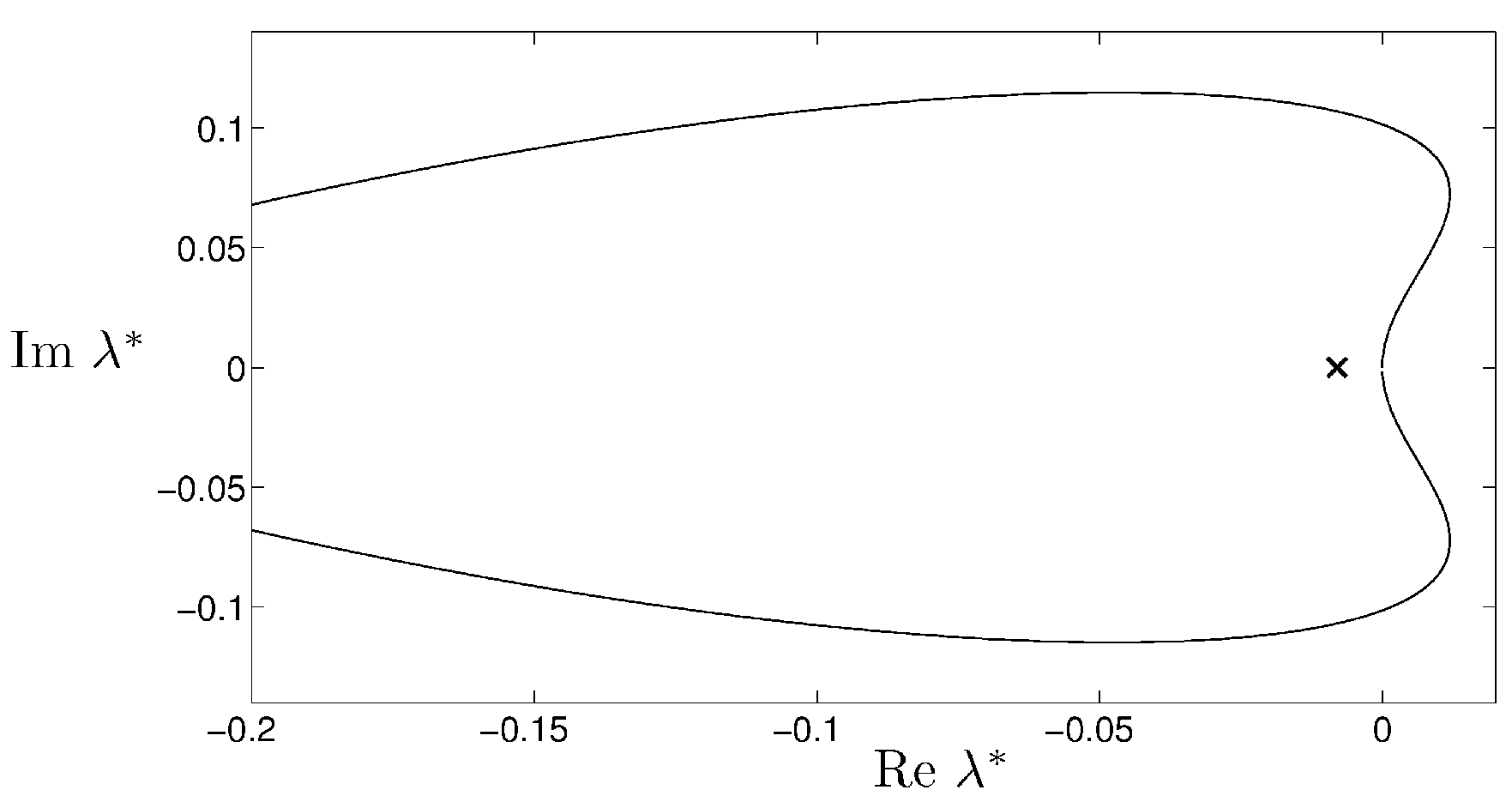}
\caption{Spectrum of $\Lia$ on an infinite domain for $\delta=0.98$
with the physical parameters corresponding W. The solid line is the
locus of the essential spectrum and the cross represents the point
spectrum.}\label{Fig:SPC A}
\end{figure}
In addition, it is straightforward to see that on a periodic domain the zero eigenfunction is
a constant:
\begin{equation}
\Pso= \binom{0}{m},
\end{equation}
so that $\Lia\Pso=0$. As shown in Fig.~\ref{Fig:f2}($b$), on an
infinite domain we have $m\to 0$ and therefore there is no such a
function in the null space of the adjoint operator that also belongs
to $L_{\mathbb{C}}^{2}$. We can therefore conclude that zero is not
in the point spectrum of $\Lia$.
The generalized zero eigenfunction on a periodic domain, i.e. the function satisfying
$\Lia\Pst=\Pso$, is found numerically and its components are shown in
Fig.~\ref{Fig:psi2}.

We can also show that the adjoint operator $\Jia$ is:
\begin{equation}
 \Jia=\begin{pmatrix}
  \mathcal{J}_i^{*1} & 0 \\
   \mathcal{J}_i^{*2} & 0
\end{pmatrix}
\end{equation}
with components:
\begin{eqnarray*}
\mathcal{J}_i^{*1} & = & \frac{18q_i}{7h_i^2}H_{ix}+\frac{17}{7h_i}\bigg(\frac{Q_i H_{ix}}{h_i}+
Q_i\partial_x\bigg)\nonumber  {} \\
& &+\frac{\eta}{2\delta}\bigg(\frac{17}{h_i^2}H_{ix}^2-
\frac{21}{h_i}H_{ixx}+\frac{9}{h_i}H_{ix}\partial_x\bigg),
\nonumber  {} \\
\mathcal{J}_i^{*2} & = & \frac{5}{\delta h_i^4}(q_iH_i+Q_i)-
\frac{1}{7 h_i^2}\bigg(q_i Q_{ix}+\frac{H_{ix}}{h_i}\bigg)\nonumber {} \\
&&-\bigg(\frac{5}{2\delta}H_{ixx}+\frac{9q_i^2-1}{7h_i^2}\bigg) \partial_x
-\frac{5}{2\delta}H_{ix}\partial_{xx} {} \\
&& -\frac{5}{6\delta}H_i\partial_{xxx}\nonumber +\frac{\eta}{\delta}\bigg[\frac{4}{h_i^2}\bigg(Q_{ix}H_{ix}-\frac{q_i H_{ix}^2}{h_i}+
\frac{H_{ix}^2}{h_i} {} \\
&& +q_i H_{ixx}-\frac{H_{ixx}}{2}-\frac34 h_i Q_{ix}\bigg)+ \bigg(\frac{4Q_i H_{ix}}{h_i^2}+ {} \\
&& \frac83\frac{H_{ix}}{h_i^2}-\frac{15}{2}\frac{Q_{ix}}{h_i}\bigg)\partial_{x}
-\frac{6}{h_i}Q_i\partial_{xx}\bigg]. \nonumber
\end{eqnarray*}

%
\end{document}